\documentclass[aps,prd,showpacs]{revtex4}
\input epsf.sty
\usepackage[dvips]{graphicx}
\def\beq{\begin{eqnarray}}
\def\eeq{\end{eqnarray}}

\begin{document} 

\newcommand{\ep}{\epsilon}
\newcommand{\fr}{\frac}
\newcommand{\reals}{\mbox{${\rm I\!R }$}}
\newcommand{\nats}{\mbox{${\rm I\!N }$}}
\newcommand{\intgs}{\mbox{${\rm Z\!\!Z }$}}
\newcommand{\cam}{{\cal M}}
\newcommand{\caz}{{\cal Z}}
\newcommand{\cao}{{\cal O}}
\newcommand{\cac}{{\cal C}}
\newcommand{\aaa}{\int\limits_{mR}^{\infty}dk\,\,}
\newcommand{\bbb}{\left[\left(\frac k R\right)^2-m^2\right]^{-s}}
\newcommand{\ccc}{\frac{\partial}{\partial k}}
\newcommand{\fff}{\frac{\partial}{\partial z}}
\newcommand{\iikma}{\aaa \bbb \ccc}
\newcommand{\ddd}{\int\limits_{mR/\nu}^{\infty}dz\,\,}
\newcommand{\eee}{\left[\left(\frac{z\nu} R\right)^2-m^2\right]^{-s}}
\newcommand{\lll}{\frac{(-1)^j}{j!}}
\newcommand{\iinma}{\ddd\eee\fff}
\newcommand{\cah}{{\cal H}}
\newcommand{\nn}{\nonumber}
\renewcommand{\theequation}{\mbox{\arabic{section}.\arabic{equation}}}
\newcommand{\komplex}{\mbox{${\rm I\!\!\!C }$}}
\newcommand{\sip}{\frac{\sin (\pi s)}{\pi}}
\newcommand{\numr}{\left(\frac{\nu}{mR}\right)^2}
\newcommand{\mzs}{m^{-2s}}
\newcommand{\rzs}{R^{2s}}
\newcommand{\abl}{\partial}
\newcommand{\g}{\Gamma\left(}
\newcommand{\ikma}{\int\limits_{\gamma}\frac{dk}{2\pi i}\,\,(k^2+m^2)^{-s}\frac{\partial}{\partial k}}
\newcommand{\ead}{e_{\alpha}(D)}
\newcommand{\sual}{\sum_{\alpha =1}^{D-2}}
\newcommand{\sulnu}{\sum_{l=0}^{\infty}}
\newcommand{\sujnu}{\sum_{j=0}^{\infty}}
\newcommand{\suani}{\sum_{a=0}^i}
\newcommand{\suanzi}{\sum_{a=0}^{2i}}
\newcommand{\zend}{\zeta_D^{\nu}}
\newcommand{\amed}{A_{-1}^{\nu ,D}(s)}
\renewcommand{\and}{A_{0}^{\nu ,D}(s)}
\newcommand{\aid}{A_{i}^{\nu ,D}(s)}
\newcommand{\res}{{\rm res}\,\,\,}
\newcommand{\sn}{\frac{\sin \pi s}{\pi}}
\newcommand{\ha}{\frac{3}{2}}
\newcommand{\smj}{\sum_{j=\ha}^\infty d(j)}
\newcommand{\smN}{\sum_{k=1}^N}
\newcommand{\smk}{\sum_{k=0}^\infty \frac{(-1)^k}{k!}}
\newcommand{\sma}{\sum_{a=0}^{2i} x_{i,a}}
\newcommand{\itz}{\int_{\frac{mR}j}^\infty dz}
\newcommand{\paran}{\left[\left(\frac{zj}R \right)^2-m^2\right]}
\newcommand{\dz}{\frac{\partial }{\partial z}}
\newcommand{\bee}{\begin{equation}}
\newcommand{\bea}{\begin{eqnarray}}
\newcommand{\eea}{\end{eqnarray}}
\newcommand{\app}[1]{\section{#1}\renewcommand{\theequation}
        {\mbox{\Alph{section}.\arabic{equation}}}\setcounter{equation}{0}}
\def\beq{\begin{eqnarray}}
\def\eeq{\end{eqnarray}}


\newcommand{\De}{\Delta}
\newcommand{\Om}{\Omega}
\newcommand{\Si}{\Sigma}
\newcommand{\G}{\Gamma}
\newcommand{\Ga}{\Gamma}
\newcommand{\Gb}{\Gamma^b}
\newcommand{\La}{\Lambda}
\newcommand{\Th}{\Theta}

\hspace{-10mm} 
\vspace*{-10.0mm} 
\thispagestyle{empty} 
{\baselineskip-4pt 
\font\yitp=cmmib10 scaled\magstep2 
\font\elevenmib=cmmib10 scaled\magstep1  \skewchar\elevenmib='177 

\vskip20mm 
\begin{center}{\large \bf 
Neutrino Dark Energy and Moduli Stabilization in a BPS Braneworld Scenario}
\end{center} 
\vspace*{4mm} 
\centerline{Andrea Zanzi} 
\vspace*{2mm}

\centerline{{\it Dipartimento di Fisica, Universit\`a di Padova,  via Marzolo 8, I-35131, Padova, Italy}}
\centerline{{\it INFN, Sezione di Padova, via Marzolo 8, I-35131, Padova, Italy}}

\centerline{\rm e--mail: zanzi@pd.infn.it}

\vspace*{4mm}

\begin{abstract}

A braneworld model for neutrino Dark Energy (DE) is presented. We consider a five dimensional two-branes set up
with a bulk scalar field motivated by supergravity. Its low-energy effective theory is derived with a moduli space approximation
(MSA). The position of the two branes are parameterized by two scalar degrees of freedom (moduli). After detuning
the brane tensions a classical potential for the moduli is generated. This potential is unstable for $dS_4$ branes
and we suggest to consider as a stabilizing contribution the Casimir energy of bulk fields. In particular we 
add a massive spinor (neutrino) field in the bulk and then evaluate the Casimir contribution
of the bulk neutrino with the help of zeta function regularization techniques. 
We construct an explicit form of the 4D neutrino mass as function of the two moduli. To recover the correct 
DE scale for the moduli potential the usual cosmological constant fine-tuning
is necessary, but, once accepted, 
this model suggests a stronger connection 
between DE and neutrino physics.

\vspace*{8mm} 
\hspace{111mm} DFPD/06/TH/04

\end{abstract}
\maketitle
\section{Introduction}

There is an increasing observational evidence that our Universe is in a state of accelerated expansion \cite{1}.
The mixing of data coming from CMB, supernovae Ia and abundance measurements illustrate a surprising 
picture of the cosmic energy budget: the dominant energy component (dark energy, DE, for reviews see \cite{quint-review}) has negative enough pressure and its 
contribution is comparable, at present, with that of the pressureless dark matter component (DM), although 
their dynamical evolution 
during the cosmological history can be totally different ("coincidence problem"). To remove, at least 
partially, this problem, a direct interaction between DM and DE has been discussed
\cite{4,5,5a,5b,amen} and the "old" idea of varying-mass particles \cite{vamps} has been taken into account again. 
These models are also well-motivated from theoretical point of view by the identification
of the DE with the string-theory dilaton (\cite{6,PT,7}, but also \cite{gaspstab} for DE as a stabilized dilaton). These 
so-called "coupled DE models"  can show 
a late-time attractor solution, i.e. the late-time cosmology is insensitive to the initial conditions for DE.
For a phenomenologically acceptable scenario, in these models the present coupling of the 
scalar field with baryons must be sufficiently weak 
while the couplings with DM must be sufficiently strong (baryo-phobic quintessence). This is a very general 
problem: allowing a direct interaction between matter and a quintessence (ultra-light $m_{\phi} \sim 10^{-33}$eV)
field could be phenomenologically dangerous (violations of the equivalence principle, time dependence of 
couplings - for reviews see \cite{uzan}). 

A possible way-out of these problems is guaranteed by 
scalar-tensor (ST) theories of gravitation \cite{bd}, where by construction matter has a
purely metric coupling with gravity (for reviews see \cite{maeda},\cite{damour},\cite{will}). The existence of an attraction
mechanism of ST towards Einstein's General Relativity (GR) has been discussed in \cite{dam3a, dam3b, max}. Simultaneously 
the presence of a second attractor is possible (\cite{max}) and the correct evolution of $\rho_{DE}$ along a so-called `tracking'
\cite{tracker} solution is rescued. However deviations from standard cosmology and General Relativity
may be present at nucleosynthesis, CMB and solar system tests of gravity \cite{catena,caffe2}. Another possible way-out is to 
consider "chameleon fields" (\cite{justin}): the mass of the scalar fields is determined by the density of the environment. On cosmological distances,
where the densities are very tiny, the mass can be of the order of the Hubble constant and the fields can roll on cosmological time scales.
On the Earth instead, the density is much higher and the field acquires a large enough mass to evade experimental bounds mantaining relevant coupling with matter. 

Recently Fardon, Nelson and Weiner (FNW) exploited the notion of "tracking" in a completely new way (\cite{fnw}): they 
suggested that DE tracks neutrinos. They established a (not so!) firm connection between the 
dark energy scale and the neutrino mass scale: 
the similarity between the two scales becomes a true relation.
The authors imagined furthermore that DE and neutrinos are directly coupled: a new dark sector arises, taking contributions
from a scalar field $\phi$ (a sort of quintessence) and neutrinos. In this scenario the dark energy density depends 
on the neutrino masses, which are not 
fixed but varying as functions of the neutrino density. The coupling between DE and neutrinos is encoded in the variable neutrino
 mass $m_{\nu}(\phi)$. During the cosmological expansion the temperature $T$ decreases. 
When $T< m_{\nu}$ neutrinos become non-relativistic and their contribution to the energy density becomes mass-dependent.
In this way the coupling $\phi-\nu$ brings to an effective potential of the form:
\begin{equation}\label{prima}
 V_{eff}=V_{0}(\phi)+n_{\nu} m_{\nu}(\phi),
\end{equation}
where $V_0$ is the scalar field potential and $n_{\nu}$ is the neutrino number density. If we assume that $V_0$ is minimized for large $m_{\nu}$,
a competition between the terms in (\ref{prima}) is created, a minimum in $V_{eff}$ arises and the neutrino mass is determined 
{\it dynamically} during
the cosmological expansion. For works on mass varying neutrinos see also \cite{peccei}, \cite{neut}.

Several points are unsatisfactory about this approach (see also \cite{peccei}): 
\begin{itemize}
 \item The FNW model is characterized by $V_{0}(\phi)$ (the potential for the scalar field) and $m(\phi)$ (the neutrino mass).
 The explicit form of these functions is inserted by-hand. It would be interesting to obtain a firmer 
theoretical derivation of these functions in a "top-down" scenario.
 \item The connection between dark energy and neutrino mass is not well established unless the scalar field gets an appropriate
value. 
\end{itemize}
As we will show, an extradimensional scenario addresses 
both drawbacks of the FNW model: \\
1) a 4D neutrino mass as a function of two scalar 
degrees of freedom (moduli) and the potential for these scalar fields are not inserted by hand but rather explicitly
calculated.\\
2) A firmer relationship is established between neutrino mass scale and Dark Energy scale.

It must be stressed that solving these problems led us to a model that is significantly different:
the only dynamic that we contemplate for the scalar fields is a chameleontic behaviour for one of the two moduli (the radion). 
Then, we have to distinguish between different distance scales: on cosmological distances, the densities are tiny and the 
radion can roll on cosmological time scales (the neutrino mass is {\it variable} with time and possible deviations
from a pure cosmological constant contribution can be observed in the DE sector); on smaller scales instead, 
the chameleon acquires a large mass (radion stabilization)
and {\it both} moduli are simply sitting on the minimum of the potential (for a discussion of possible manifestations 
of DE at short scales
see \cite{caffe}).
As a final result, the connection between DE and neutrino mass will be established in the form
\beq
(\frac{V^{1/4}}{m_{\nu}})_{min}\sim 1,
\eeq
where $V$ is the total moduli potential. 

Before getting into the detailed presentation of the model, 
here is a summary of our approach. We consider 
a supersymmetric extension of Randall-Sundrum two branes model (\cite{RS1,RS2}, while 
for reviews on extra-dimensions see \cite{extra}).
This kind of approach finds its theoretical firmer justification in 
string theory (see \cite{HW,witten}).  
Our general set-up will follow \cite{brax,brax1}: a five-dimensional braneworld model 
with two boundary branes and a scalar field in the bulk motivated
by supergravity.
As  we will see in the forecoming sections, the most
distinguishing feature of this (BPS) class of models is that the brane and bulk dynamics are related to each 
other and under control. Once we have solved bulk equations (Einstein equations and Klein-Gordon equation) the 
boundary conditions do not
give any additional information, so we can put the branes without any obstacle into the background. The positions of the 
two branes are specified by two independent constants (no-force condition; for a discussion on BPS condition and connections between 
brane physics and black-hole physics see  \cite{gionni}). 
We will discuss than the moduli space approximation to derive 
the low energy action and a bi-scalar-tensor theory is recovered, 
with the two scalar fields $\lambda$ and $\phi$ (moduli) parametrizing the positions of the branes: 
the two constants are promoted to scalar degrees
of freedom. To render this model more realistic, 
we can leave the BPS configuration in two different ways: i) putting matter on the branes; ii) perturbing 
the brane tensions (i.e. parametrizing the supersymmetry breaking in a phenomenological way). It has been 
shown (\cite{gonz}) that it is 
possible to stabilize both moduli putting matter on branes without perturbing the brane tensions. However, since we want 
to describe a Universe with a positive cosmological constant, the detuning of the brane tensions is unavoidable and a 
classical potential for the moduli is generated (\cite{baggerredi, marsiglia}).
Unfortunately the potential is unstable and, to stabilize the system, we suggest to consider as a stabilization
mechanism the quantum (Casimir) contribution 
from bulk fields. We will then add a massive spinor (neutrino) in the bulk and we will evaluate the Casimir 
energy using zeta function regularization techniques (for reviews on Casimir energy see \cite{casimir}, \cite{kirsten};
for zeta function techniques see \cite{TEN}, \cite{kirsten}, \cite{MOSS2}). Remarkably the potential
$V=V(\lambda,\phi)$ is not inserted by-hand in our model and the geometrical configuration of the system is stabilized 
(a long-standing problem in extra-dimensional 
and string-inspired scenarios, see for example \cite{buchbinder} and
references therein for a general treatment, but also \cite{gonz, webster} for moduli stabilization in BPS braneworlds).

As far as the 4d neutrino mass is concerned, following \cite{grossman}, we will obtain a small 4d neutrino mass
without recurring to a see-saw mechanism.

About the organization of this article, in section 2 we present the action and the moduli space approximation; in section 3 we study
the chameleon mechanism for the radion; in section 4 the explicit form of the neutrino mass is derived as function of the moduli;
in section 5 the moduli potential is analyzed. In section 6 we will discuss the parameters of the model and the moduli stabilization. In section 7 we will draw some concluding remarks. The appendix contains some technical details.

\section{The model}

The bulk action is the sum of two main contributions: the 
first one (SUGRA) is motivated by supergravity and follows \cite{brax, brax1}, the
second one (SPINOR) is the bulk-neutrino part. We have

\begin{equation}
S_{\rm bulk} = S_{\rm SUGRA} + S_{\rm SPINOR}
\end{equation}

where the SUGRA action consists of two terms which describe 
gravity and the bulk scalar field ($C$) dynamics:
\begin{equation}
S_{\rm SUGRA} = \frac{1}{2\kappa_5^2} \int d^5 x \sqrt{-g_5}
\left( {\cal R} - \frac{3}{4}\left( (\partial C)^2 + U \right)\right);
\end{equation}

while the SPINOR action, which takes into account a Dirac bulk fermion with mass $m$ 
of order the fundamental scale $M$, is written in the form: 

\begin{equation}
\label{action}
   S = \int\!\mbox{d}^4x\!\int\!\mbox{d}z\,\sqrt{-g_5}
   \left\{ E_a^A \left[ \frac{i}{2}\,\bar\Psi\gamma^a
   (\partial_A-\overleftarrow{\partial_A})\Psi
   + \frac{\omega_{bcA}}{8}\,\bar\Psi \{\gamma^a,\sigma^{bc}\} \Psi
   \right] - m\,\mbox{sgn}(\alpha)\,\bar\Psi\Psi \right\} \,.
\end{equation}
We use capital
indices $A,B,\dots$ for objects defined in curved space, and 
lower-case indices $a,b,\dots$ for objects defined in the tangent 
frame. The matrices $\gamma^a=(\gamma^\mu,i\gamma_5)$ provide a 
four-dimensional representation of the Dirac matrices in 
five-dimensional flat space. The quantity
$E_a^A$ is the
inverse vielbein and $\omega_{bcA}$ is the spin connection. Because 
in our case the metric is diagonal the connection gives no contribution to
the action (2.5).

The sign change of the mass term under $\alpha\to-\alpha$ is necessary in 
order to conserve $\alpha$-parity, as required by the $Z_2$ orbifold 
symmetry we impose.

Further, our setup contains two branes. One of these branes has a
positive tension, the other brane has a negative tension.  They are
described by the action
\begin{eqnarray}
S_{\rm brane 1} &=& -\frac{3}{2\kappa_5^2}\int d^5x \sqrt{-g_5} U_B
\delta(z_1), \label{b1} \\
S_{\rm brane 2} &=& +\frac{3}{2\kappa_5^2}\int d^5x \sqrt{-g_5} U_B \delta(z_2) \label{b2}.
\end{eqnarray}
In these expressions, $z_1$ and $z_2$ are the (arbitrary) positions of
the two branes, $U_B$ is the superpotential; $U$, the bulk potential
energy of the scalar field, is given by (BPS relation)
\begin{equation}
U = \left(\frac{\partial U_B}{\partial C}\right)^2 - U_B^2.
\end{equation}
This specific configuration, when no fields other than the bulk scalar field are present and when the
bulk and brane potentials are unperturbed (no susy breaking effects), is the BPS configuration (\cite{gonz}). When $U_B$ 
is the constant potential, the Randall-Sundrum model is recovered with a bulk cosmological constant
$ \Lambda_5=(3/8) U=-(3/8) U_B^2$.

We will also include the Gibbons--Hawking boundary term for each
brane, which have the form
\begin{equation}
S_{\rm GH} = \frac{1}{\kappa_5^2}\int d^4 x \sqrt{-g_4} K,
\end{equation}
where $K$ is the extrinsic curvature of the individual branes.

The solution of the system above can be derived from 
BPS--like equations of the form
\begin{equation}
\frac{a'}{a}=-\frac{U_B}{4},\ C'=\frac{\partial U_B}{\partial C},
\end{equation}
where $'=d/dz$ for a metric of the form
\begin{equation}\label{background}
ds^2 = dz^2 + a^2(z)\eta_{\mu\nu}dx^\mu dx^\nu.
\end{equation}
We will particularly focus on the case where the superpotential is an 
exponential function:
\begin{equation}\label{potential}
U_B=4k e^{\alpha C}.
\end{equation}
The solution for the scale factor reads
\begin{equation}\label{scale}
a(z)=(1-4k\alpha^2z)^{\frac{1}{4\alpha^2}},
\end{equation}
while the scalar field solution is
\begin{equation}\label{psi}
C = -\frac{1}{\alpha}\ln\left(1-4k\alpha^2z\right).
\end{equation}
In the $\alpha\to 0$ we retrieve the AdS profile
\begin{equation}
a(z)=e^{-kz}.
\end{equation}
In that case the scalar field decouples and the singular point (for which the scale factor vanishes) 
at $z_* = 1/(4k\alpha^2)$ 
is removed (singularities in braneworld scenarios are analyzed in \cite{naked}).

In the following we will discuss the moduli space approximation. Two of the moduli of the
system are the brane positions.  That is, in the solution above the
brane positions are arbitrary.  In the moduli space approximation,
these moduli are assumed to be space-time dependent since in connection to matter distribution on branes.  We denote the
position of ("visible") brane 1 with $z_1 = \phi(x^\nu)$ and the position of ("hidden")
brane 2 with $z_2 = \lambda(x^\mu)$. We consider the case where the
evolution of the brane is slow. This means that in constructing the
effective four--dimensional theory we neglect terms like $(\partial
\phi)^3$.

In addition to the brane positions, we need to include the graviton 
zero mode, which can be done by replacing $\eta_{\mu\nu}$ with a 
space--time dependent tensor $g_{\mu\nu}(x^\mu)$. Note that the moduli 
space approximation is only a good approximation
if the time--variation of the moduli fields is small. This should be
the case for late--time cosmology well after nucleosynthesis, which we 
are interested in.

\subsection{Moduli Space Approximation: the gravitational sector}
Let us first consider the SUGRA action. Replacing $\eta_{\mu\nu}$ 
with $g_{\mu\nu}(x^{\mu})$ in (\ref{background}) we have for the 
Ricci scalar ${\cal R} = {\cal R}^{(4)}/a^2 + \tilde {\cal R}$, where $\tilde {\cal R}$ is the 
Ricci--scalar of the background (\ref{background}). We explicitly use
the background solution (\ref{scale}) and (\ref{psi}), so that 
$\tilde R$ will not contribute to the low--energy effective action. 
Also, in this coordinate system, where the branes move, there is no 
contribution from the part of the bulk scalar field. Collecting 
everything we therefore have
\begin{equation}
S_{\rm bulk} = \frac{1}{2\kappa_5^2} \int dz d^4 x a^4
\sqrt{-g_4}\frac{1}{a^2}{\cal R}^{(4)} = \int d^4 x 
\sqrt{-g_4} f(\phi,\lambda) {\cal R}^{(4)},
\end{equation}
with 
\begin{equation}
f(\phi,\lambda) = \frac{1}{2 \kappa_5^2} \int^{\lambda}_{\phi} dz a^2 (z).
\end{equation}
We remind the reader that $a(z)$ is given by (\ref{scale}). 

We now turn to the boundary terms. The integrals 
(\ref{b1}) and (\ref{b2}) do not contribute to the effective action 
for the same reason that $\tilde R$ does not contribute. Let us 
therefore turn our attention to the Gibbons--Hawking boundary terms.

First, it is not difficult to construct the normal vectors to the
brane:
\begin{equation}
n^\mu = \frac{1}{\sqrt{1+ (\partial \phi)^2/a^2}}\left( \partial^\mu
\phi/a^2,1\right).
\end{equation}

Then the induced metric on each brane is given by
\begin{eqnarray}
g_{\mu\nu}^{\rm ind,1} &=& a^2(\phi) g_{\mu\nu}^4 - \partial_{\mu}\phi
\partial_{\nu} \phi, \\
g_{\mu\nu}^{\rm ind,2} &=& a^2(\lambda) g_{\mu\nu}^4 - \partial_{\mu}\lambda
\partial_{\nu} \lambda.
\end{eqnarray}
Thus, 
\begin{eqnarray}
\sqrt{-g^{\rm ind,1}} &=& a^4(\phi) \sqrt{-g_4}\left[ 1 -
\frac{1}{2a^2(\phi)}(\partial \phi)^2\right], \\
\sqrt{-g^{\rm ind,2}} &=& a^4(\lambda) \sqrt{-g_4}\left[ 1 -
\frac{1}{2a^2(\lambda)}(\partial \lambda)^2\right].
\end{eqnarray}
So the Gibbons--Hawking boundary terms take the form 
\begin{eqnarray}
\frac{1}{\kappa_5^2} \int d^4 x a^4 \sqrt{-g_4}\left[ 1 -
\frac{1}{2a^2(\phi)} (\partial \phi)^2 \right] K, \\
\frac{1}{\kappa_5^2} \int d^4 x a^4 \sqrt{-g_4}\left[ 1 -
\frac{1}{2a^2(\lambda)} (\partial \lambda)^2 \right] K.
\end{eqnarray}
The trace of the extrinsic curvature tensor can be calculated from 
\begin{equation}
K = \frac{1}{\sqrt{-g_5}}\partial_\mu \left[ \sqrt{-g_5}n^\mu \right].
\end{equation}
Neglecting higher order terms this gives
\begin{eqnarray}
K = 4\frac{a'}{a}\left[1 - \frac{(\partial \phi)^2}{4a^2}\right].
\end{eqnarray}
The terms for the second brane can be obtained analogously.
Using the BPS conditions and keeping only the kinetic terms, we get
for the Gibbons--Hawking boundary terms
\begin{eqnarray}
&+&\frac{3}{4\kappa_5^2}\int d^4x \sqrt{-g_4} a^2(\phi) U_B(\phi)
(\partial \phi)^2, \\
&-&\frac{3}{4\kappa_5^2}\int d^4x \sqrt{-g_4} a^2(\lambda) U_B(\lambda)
(\partial \sigma)^2.
\end{eqnarray}
Collecting all terms we find
\begin{eqnarray}
S_{\rm MSA} = \int d^4 x \sqrt{-g_4}\left[ f(\phi,\lambda) {\cal R}^{(4)} 
+ \frac{3}{4}a^2(\phi)\frac{U_B(\phi)}{\kappa_5^2}(\partial \phi)^2 
- \frac{3}{4} a^2(\lambda)\frac{U_B}{\kappa_5^2}(\lambda)(\partial \lambda)^2 \right].
\end{eqnarray}
Note that the kinetic term of the field $\phi$ has the wrong
sign. This is an artifact of the frame we use here. It is possible to go to the Einstein frame with a simple
conformal transformation, in which the sign in front of the kinetic
term is correct for both fields. We observe that for a BPS system the moduli potential is totally absent: the fields describe flat directions.

Coming back to the action above, we redefine the fields in the following way:
\begin{eqnarray}
\tilde \phi^2 &=& \left(1 - 4k\alpha^2 \phi\right)^{2\beta}, \label{posia1}\\
\tilde \lambda^2 &=& \left(1-4k\alpha^2 \lambda\right)^{2\beta} \label{posia2},
\end{eqnarray}
with 
\begin{equation}
\beta = \frac{2\alpha^2 + 1}{4\alpha^2};
\end{equation}
then, the gravitational sector can be written as
\begin{eqnarray}
S_{\rm MSA} &=&\frac{1}{2k\kappa_5^2(2\alpha^2 + 1)}\int d^4 x \sqrt{-g_4}\left[ \left(\tilde\phi^2 -
\tilde\lambda^2 \right) {\cal R}^{(4)} +
\frac{6}{2\alpha^2 + 1}\left( (\partial \tilde\phi)^2
-(\partial \tilde\lambda)^2\right)\right].
\end{eqnarray}

This is an action of the form of a multi--scalar tensor theory, in
which one scalar field has the wrong sign in front of the kinetic
term. Furthermore, in this frame there is a peculiar point where the
factor in front of ${\cal R}$ can vanish, namely when
$\tilde\phi=\tilde\lambda$, which corresponds to colliding branes. 

In order to avoid mixed
terms like $(\partial_\mu \tilde\phi)(\partial^\mu \tilde\lambda)$, we
shall define two new fields\footnote{Do not confuse the Ricci scalar
${\cal R}$ with the new field $R$.}:
\begin{eqnarray}
\tilde \phi &=& Q \cosh R, \label{posib1} \\
\tilde \lambda &=& Q \sinh R \label{posib2}.
\end{eqnarray}
To go to the Einstein frame we perform a conformal transformation:
\begin{equation}
\tilde g_{\mu\nu} = Q^2 g_{\mu\nu}.
\end{equation}
Then 
\begin{equation}
\sqrt{-g} Q^2 {\cal R} = \sqrt{-\tilde g} \left( \tilde {\cal R} -
\frac{6}{Q^2}(\tilde\partial Q)^2 \right) .
\end{equation}
Collecting everything we get the action in the Einstein frame
(where we now drop the tilde):
\begin{eqnarray}
S_{\rm EF} &=& \frac{1}{2k\kappa^2_5(2\alpha^2 + 1)} 
\int d^4x \sqrt{-g}\left[ {\cal R} -  \frac{12\alpha^2}{1+2\alpha^2}
\frac{(\partial Q)^2}{Q^2} - \frac{6}{2\alpha^2 + 1}(\partial R)^2\right].
\end{eqnarray}
Clearly, in this frame both fields have the correct sign in front of
the kinetic terms. Note that for $\alpha \rightarrow 0$ 
(i.e.\ the Randall--Sundrum case) the $Q$--field decouples. 
In this case, the field $R$ plays the role of the radion, i.e. it 
measures the distance between the branes. Furthermore, we can identify 
the gravitational constant:
\begin{equation}
16\pi G = 2k\kappa_5^2 (1+2\alpha^2).
\end{equation}

\subsection{Moduli space approximation: the matter sector}
In this section we are going to leave the BPS configuration. Let's suppose we put some matter on the branes:
Standard Model fields on one brane ("visible") and Dark Matter particles on the other brane ("hidden").
When the cosmological evolution of the branes is analyzed in the absence 
of supersymmetry breaking potentials, it has been shown (\cite{gonz}) that the branes are driven by their matter
content: one brane is driven towards the minimum of $U_B$, while the other brane towards the maximum. 
In this way it is possible to achieve a full stabilization of the system, granted that we choose the superpotential
of the theory in a proper way. If we don't localize matter on branes and don't break supersymmetry, than the BPS
configuration is respected, the no-force condition between the branes is present and the system is static.

However it is our intention to describe a Universe with a positive cosmological constant and detuning is necessary in order to
achieve de Sitter geometry. We therefore introduce matter as well as supersymmetry breaking 
potentials $V(Q,R)$ and $W(Q,R)$ on each branes. We begin with the potentials:
to first order in the moduli space approximation we get
\begin{equation}
\int d^4 x \sqrt{-g_4} \left[ a^4(\phi)V(\phi) \right]
\end{equation}
with $a^4(\phi) = \tilde \phi^{4/(1+2\alpha^2)}$. The expression for 
a  potential $W$ on the second brane is similar with 
$a(\phi)$ is replaced by $a(\lambda)$. In the Einstein frame we have
(dropping the tilde from the metric):
\begin{equation}
\int d^4 x\sqrt{-g} Q^{-8\alpha^2/(1+2\alpha^2)}(\cosh
R)^{4/(1+2\alpha^2)} V(Q,R) \equiv \int d^4 x\sqrt{-g} V_{\rm eff}(Q,R),
\end{equation}
where we have defined 
\begin{equation}
V_{\rm eff}(Q,R) = Q^{-8\alpha^2/(1+2\alpha^2)}(\cosh
R)^{4/(1+2\alpha^2)} V(Q,R).
\end{equation}
The expression for $W(Q,R)$ in the Einstein frame is
\begin{equation}
\int d^4 x\sqrt{-g} Q^{-8\alpha^2/(1+2\alpha^2)}(\sinh
R)^{4/(1+2\alpha^2)} W(Q,R) \equiv \int d^4 x\sqrt{-g} W_{\rm eff}(Q,R),
\end{equation}
where
\begin{equation}
W_{\rm eff}(Q,R) = Q^{-8\alpha^2/(1+2\alpha^2)}(\sinh
R)^{4/(1+2\alpha^2)} W(Q,R).
\end{equation}
For matter, the action has the form
\begin{equation}
S_m^{(1)} = S_m^{(1)}(\Psi_1,g^{ind,1}_{\mu\nu}) \hspace{0.5cm} {\rm and}
\hspace{0.5cm} S_m^{(2)} = S_m^{(2)}(\Psi_2,g^{ind,2}_{\mu\nu}),
\end{equation}
where $g^{ind}$ denotes the {\it induced} metric on each branes and
$\Psi_i$ the matter fields on each branes. Note that we do not 
couple the matter fields $\Psi_i$ to the bulk scalar field, and thus 
not to the fields $Q$ and $R$. In going to the Einstein frame we get 
\begin{equation}
S_m^{(1)} = S_m^{(1)}(\Psi_1,A^2(Q,R)g_{\mu\nu}) \hspace{0.5cm} {\rm and}
\hspace{0.5cm} S_m^{(2)} = S_m^{(2)}(\Psi_2,B^2(Q,R)g_{\mu\nu}),
\end{equation}
In this expression we have used the fact that, in going to the
Einstein frame, the induced metrics on each branes transform with a
different conformal factor, which we denote with $A$ and $B$. We have neglected the derivative terms in the moduli fields when considering the coupling to matter on the brane. They lead to higher order operators which can be easily incorporated.  
The energy--momentum tensor in the Einstein frame is defined as
\begin{equation}
T_{\mu\nu}^{(1)} = 2 \frac{1}{\sqrt{-g}} \frac{\delta
S_m^{(1)}(\Psi,A^2(Q,R)g)}{\delta g^{\mu\nu}}
\end{equation}
with an analogous definition for the energy--momentum tensor for
matter on the second brane.

In the Einstein frame, the total action is therefore, 
\begin{eqnarray}
S_{\rm EF} &=& \frac{1}{16\pi G} 
\int d^4x \sqrt{-g}\left[ {\cal R} -  \frac{12\alpha^2}{1+2\alpha^2}
\frac{(\partial Q)^2}{Q^2} - \frac{6}{2\alpha^2 + 1}(\partial
R)^2\right] \nonumber \\
&-& \int d^4 x\sqrt{-g} (V_{\rm eff}(Q,R)+W_{\rm eff}(Q,R)) 
+ S_m^{(1)}(\Psi_1,A^2(Q,R)g_{\mu\nu}) + S_m^{(2)}(\Psi_2,B^2(Q,R)g_{\mu\nu}).
\end{eqnarray}

We point out that in this braneworld scenario we naturally obtain a multimetric theory. 
Particles localized in different branes "feel" the respective induced metric ($g_{\mu\nu}^{ind}$).
In the next section we will discuss an interesting phenomenological application of this fact: the chameleon 
mechanism.
The moduli potential will be presented in a forecoming section. For more details on the moduli 
space approximation see \cite{gonz}.

\section{Chameleon fields}

As we already mentioned in the introduction, the coupling of the moduli with matter is severely 
constrained by phenomenological tests. Introducing massless or ultra-light scalar fields with generic 
coupling with matter is prohibited by tests on the equivalence principle and on variations of the 
fundamental constants. Different ways to avoid this conflict have been considered
in literature. Basically the fields introduced in the model have either a small coupling today or
they have a large enough mass (stabilization). Recently a new stabilization scenario has been introduced in \cite{justin}. 
According to these authors, the mass of the scalar fields is determined by the density of the environment. On cosmological distances,
where the densities are very tiny, the mass can be of the order of the Hubble constant and the fields can roll on cosmological time scales.
On the Earth instead, the density is much higher and the field acquires a large enough mass to evade all current experimental bounds
on deviations from GR. In other words, the scalar field shows a "chameleontic behaviour".
In this section we will discuss this kind of scalar fields. 
We will start by reviewing the mechanism with a basic example; then
we will discuss the role of the radion as possible chameleon field.

\subsection{The chameleon mechanism}

Let's consider a multi-metric model with the following scalar-tensor action:

\begin{equation}
S =\int d^4 x\sqrt{-g} \left\{\frac{M_{Pl}^2R}{2} - \frac{(\partial \phi)^2}{2} - 
 V(\phi) + {\cal L}_m(\psi_m^{i},g^{(i)}_{\mu\nu})\right\}
\end{equation}
where $\phi$ is a generic scalar field with potential $V(\phi)$, $\psi_m^{(i)}$ denotes
matter fields which follow geodesics of a metric $g^{(i)}_{\mu\nu}=A^2_i(\phi)g_{\mu\nu}$. The scalar field interacts with matter through the conformal factor $A^2(\phi)$, so
the dynamics of the chameleon $\phi$ is described by an effective potential that for pressureless (non-relativistic)
matter takes the form (with general multi-metric theory a sum over $i$ is present)

\begin{equation}
V_{\rm eff}(\phi)=V(\phi) +\rho_m A(\phi)\,.
\end{equation}
In other words, the conformal coupling modifies the Klein-Gordon equation for the chameleon
in the following way:
\begin{equation}
\nabla^2 \phi = V_{,\phi} + \alpha_\phi\rho_m A(\phi)\,,
\label{KG2}
\end{equation}
where $\rho_m$ is conserved with respect to the Einstein frame metric ($g_{\mu\nu}$).
Supposing a run-away behaviour for the "bare" potential $V(\phi)$, the effective 
potential will show a minimum if a "competition" between $V(\phi)$ and the density-dependent term is
realized (i.e. $A(\phi)$ is an increasing function).  In this way the chameleontic properties of the field are recovered:
the location of the minimum and the mass of the fluctuactions $m^2=V_{eff}''$ both depend on $\rho_m$. 

Another effect which suppresses the chameleon-mediated force is the following: for large enough objects, the $\phi$-force 
on a test particle is almost totally due to a thin shell just below the surface of the object, while the contribution of core-matter
is negligible. Only a small fraction of the total mass affects the motion of a test particle outside.
This mechanism has been named "thin-shell mechanism". To give a more detailed derivation, we consider a chameleon solution
for a spherically symmetric object of radius R and homogeneous density $\rho$.
We take $V(\phi) = M^{4+n}/\phi^n$, where $M$ has units of mass, and an exponential coupling of the form $A(\phi)= e^{\beta\phi}$ with
$\beta={\cal O}(1)$. Next we choose boundary conditions requiring that the solution be non-singular at the origin and that the 
chameleon tends to its environment value $\phi_0$ far from the object.

For sufficiently large objects the field assumes inside the object a value $\phi_c$ which minimizes the effective
potential, i.e. $V_{,\phi}(\phi_c) + \beta\rho_c e^{\beta\phi_c/M_{Pl}}/M_{Pl} = 0$. This holds everywhere inside the object except
within a thin shell of thickness $\Delta R$ below the surface where the field grows. Outside the object, the profile for
$\phi$ is essentially that of a massive scalar, $\phi \sim \exp(-m_0r)/r$, where $m_0$ is the mass of the chameleon in the ambient medium.

The thickness of the shell is related to $\phi_0$, $\phi_c$,
and the Newtonian potential of the object, $\Phi_N=M/8\pi M_{Pl}^2R$, by
\begin{equation}
\frac{\Delta R}{R} \approx \frac{\phi_\infty-\phi_c}{6\beta
M_{Pl}\Phi_N}\,.
\label{DR}
\end{equation}
The exterior solution can then be written explicitly as~\cite{justin}
\begin{eqnarray}
\phi(r) &\approx& -\left(\frac{\beta}{4\pi M_{Pl}}\right)\left(\frac{3\Delta
R}{R}\right)\frac{M e^{-m_0 (r-R)}}{r} + \phi_0\,. \label{thinsoln}
\end{eqnarray}

This equation gives the correction to Newton's law at short distances in the form
$F= (1+\theta) F_N$ where $\theta = 2\beta^2 \left(\frac{3\Delta R}{R}\right)$. Hence a thin-shell $\Delta R/R<<1$
guarantees a small deviation from Newton's law.

\subsection{The radion as a chameleon}
In the general set-up we described in the first section, the possible interpretation of the radion $R$ (interbrane distance) 
as a dark energy chameleon field has been investigated in \cite{radion}.
These authors introduced by-hand a run-away bare potential for the radion, to avoid the falling
of the hidden brane into the naked singularity, in the form
\begin{equation}
V(R) = \Lambda^4 R^{-\gamma}.
\end{equation}
Their investigation studied the efficiency of the thin-shell mechanism for the radion. The coupling to matter
was written by the expansion

\begin{equation}\label{approx}
A(R) \approx 1 + \frac{1}{6}\frac{R^2}{2} +...,
\end{equation}
where the neglection of higher-order terms was allowed by the cosmological evolution, that drove
the radion to small field value. Furthermore, the authors calculated a radion-mediated force, with
a correction to Newton's law given by
\begin{equation}\label{radioncoupling}
\theta=\frac{1}{18} R_\infty^2.
\end{equation}
The strength of the radion-mediated force is specified by the cosmological value of the radion.
Since $\theta$ must be small the parameters of the theory must be adjusted to guarantee a small field value.
In particular, imposing the gravitational constraints implied $\Lambda \sim 10^{-3} $ eV
and $\gamma\le 10^{-5}$. In other words the model must be extremely fine-tuned if we want simultaneously
(1) to interpret the radion as a dark energy candidate and (2) to satisfy local gravitational constraints.
The fine-tuning of $\gamma$ can be avoided by considering the potential $ V(R)= \Lambda ^4 e^{(\frac{\Lambda}{R})^n}$, but
the fine-tuning on $\Lambda$ remains as a cosmological constant fine-tuning.

We are going to exploit the chameleon mechanism to stabilize the radion (R-modulus) following the approach
of \cite{radion}. A detailed description of the moduli stabilization will be given in section 6.  

To proceed further, our first step will be the creation of a "naturally small"
neutrino mass in this braneworld set-up (see next section). The second step will be the construction of the potential
for the scalar field(s) using a physical mechanism and not simply including potentials by-hand (see section 5).

\section{The neutrino mass}

In this section we describe how the sterile bulk fermion can provide a mechanism (different from the see-saw)
to obtain a small Dirac neutrino mass in four dimensions. In our scenario all standard model matter and gauge fields are localized on the visible brane.
We begin with a brief review of the scenario with mass-varying neutrinos recently proposed in \cite{fnw}. 
We proceed with the description of the Kaluza-Klein reduction, then we show that it is possible to 
localize the sterile fermion near the hidden brane following \cite{grossman}.
In the end of the section we will discuss the role of the conformal transformation to the Einstein frame.

\subsection{Mass varying neutrinos and dark energy: the need for moduli stabilization}
 
We are now going to describe in more detail the idea proposed recently by Fardon, Nelson and Weiner \cite{fnw}
that connects the dark energy sector with neutrino's one. 

The main assumption in the FNW scenario is that the dark energy $sector$ has two components: 
the dark energy contribution of some (for example) quintessence field $\phi$ plus the neutrino energy density
\begin{equation}
\rho_{{\rm dark}}=\rho_{\nu} +\rho_{\rm{dark~ energy}}.
\end{equation}

An interaction is present between the two components, due to the dependence of the neutrino mass on
the quintessence field. The form of this dependence is obtained via a see-saw mechanism with a variable mass $M=M(\phi)$
for the right-handed neutrino. Remarkably the functional dependence $M(\phi)$ has been put in by-hand. 
It could be $M(\phi)=h\phi$ or $M(\phi)=M_0e^{\phi^2/f^2}$ with $h$ and $f$ some constants or something else. 
In the model considered in this article, however, we will be able to derive explicitly a form of the neutrino mass as function of the two
moduli. In our model variable neutrino mass is not put in by-hand.
During the expansion of the Universe, the temperature falls down till $T<m_{\nu}$.
At that time the neutrinos become non-relativistic particles and an effective potential is generated in the form
\begin{equation}
\rho_{{\rm dark}}=m_{\nu}n_{\nu} +\rho_{\rm{dark ~energy}}(m_{\nu}).
\end{equation}

FNW assumed that $\rho_{dark}$ is stationary with respect to variations in the neutrino mass. In other words 
the neutrino mass is fixed by a competition between $\rho_{dark energy}$ and $\rho_{\nu}$. The hypothesis of stationarity
implies that

\begin{equation}
\frac{\partial \rho_{\rm{dark}}}{\partial m_{\nu}}=n_{\nu} +\frac{\partial\rho_{\rm{dark ~energy}}(m_{\nu})}{\partial m_{\nu}} =0.
\end{equation}
This last equation determines the temperature dependence of the neutrino mass in a specific dark energy model.
Adding to this scheme the conservation of energy equation
\begin{equation}
\dot{\rho}=- 3H(\rho + p),
\end{equation}
it is possible to show that
\begin{equation}
\omega +1= \frac {m_{\nu}n_{\nu}}{\rho_{\rm{dark}}}= \frac{m_{\nu}n_{\nu}}{
m_{\nu}n_{\nu}+\rho_{\rm{dark ~energy}}}.
\end{equation}
where
\begin{equation}
\omega = \frac{p_{dark }}{\rho_{dark}}.
\end{equation}
Equation (61) inform us that the neutrino contribution to $\rho_{dark}$ is a small fraction
of $\rho_{dark}$.
Another important result can be obtained observing that the continuity equation guarantees $\rho_{dark} \sim R^{-3(1+w)}$.
Since $n_{\nu} \sim R^{-3}$, the neutrino mass is nearly inversely proportional to the neutrino density:

\begin{equation}
m_{\nu} \sim n_{\nu}^{\omega}\simeq 1/n_{\nu}.
\end{equation}

For further details we refer the reader to references \cite{peccei},\cite{neut}.

As pointed-out by Peccei \cite{peccei}, in the FNW scenario the intrinsic scale of dark energy is not obtained 
naturally since the value of the potential in the present epoch is fixed by hand:
\begin{equation}
V(m_{\nu}^0) \sim T^3_0 m_{\nu}^0 \sim \rho_{DE} \sim (2 \times 10^{-3} eV)^4.
\end{equation}
Thus, the connection of the dark energy scale with the neutrino mass scale is translated into a stabilization 
problem for the field $\phi$. When neutrinos become non-relativistic and $m_{\nu}n_{\nu}$ switches on, is it possible to 
guarantee the correct minimum?

Remarkably, in our approach we will obtain a neutrino mass dependent on the moduli $\lambda$ and $\phi$. 
In other words the neutrino mass depends on the geometrical configuration of the braneworld. As we already mentioned in 
the introduction, the only dynamic that we contemplate for the scalar fields is a 
chameleontic behaviour for the radion. 
On cosmological distances, 
the densities are tiny and the 
radion can roll on cosmological time scales (the neutrino mass is {\it variable} 
with time and possible deviations
from a pure cosmological constant contribution can be observed in the DE sector); 
on smaller scales instead, densities are higher and 
{\it both} moduli are simply sitting on the minimum of the potential.
As a final result, the connection between DE and neutrino mass will be established in the form
\beq
(\frac{V^{1/4}}{m_{\nu}})_{min}\sim 1,
\eeq
where $V$ is the total moduli potential.

\subsection{Kaluza-Klein reduction}

Let's consider again the SPINOR action

\begin{equation}\label{SPIN}
   S_{SPINOR} = \int\!\mbox{d}^4x\!\int\!\mbox{d}z\,\sqrt{-g_5}
   \left\{ E_a^A \left[ \frac{i}{2}\,\bar\Psi\gamma^a
   (\partial_A-\overleftarrow{\partial_A})\Psi
   + \frac{\omega_{bcA}}{8}\,\bar\Psi \{\gamma^a,\sigma^{bc}\} \Psi
   \right] - m\,\mbox{sgn}(\alpha)\,\bar\Psi\Psi \right\} \,.
\end{equation}
We use capital
indices $A,B,\dots$ for objects defined in curved space, and 
lower-case indices $a,b,\dots$ for objects defined in the tangent 
frame. The matrices $\gamma^a=(\gamma^\mu,i\gamma_5)$ provide a 
four-dimensional representation of the Dirac matrices in 
five-dimensional flat space. The quantity
$E_a^A$ is the
inverse vielbein and $\omega_{bcA}$ is the spin connection. Because 
in our case the metric is diagonal the connection gives no contribution to
the action (\ref{SPIN}).

The sign change of the mass term under $\alpha\to-\alpha$ is necessary in 
order to conserve $\alpha$-parity, as required by the $Z_2$ orbifold 
symmetry we impose.

Using an integration by parts and defining left- and right-handed
spinors $\Psi_{L,R}\equiv\frac12(1\mp\gamma_5)\Psi$, the action can 
be written as
\begin{eqnarray}\label{Sdel}
   S_{SPINOR} &=& \int\!\mbox{d}^4x\,\sqrt{-g_4}\int\!\mbox{d}z\,\bigg\{
    e^{-3\sigma} \left( \bar\Psi_L\,i\rlap/\partial\,\Psi_L
    + \bar\Psi_R\,i\rlap/\partial\,\Psi_R \right) 
    - e^{-4\sigma}\,m\,\mbox{sgn}(\alpha) \left( 
    \bar\Psi_L \Psi_R + \bar\Psi_R \Psi_L \right) \nonumber\\
   &&\mbox{}- \frac{1}{2} \left[ \bar\Psi_L \left(
    e^{-4\sigma} \partial_z + \partial_z\,e^{-4\sigma}
    \right) \Psi_R - \bar\Psi_R \left(
    e^{-4\sigma} \partial_z + \partial_z\,e^{-4\sigma}
    \right) \Psi_L \right] \bigg\} \,,
\end{eqnarray}
where we considered the redefinition $a(z)=e^{-\sigma}$ and we observe that $\sqrt{-g_5}=e^{-4\sigma}\sqrt{-g_4}$.
The action is even under the 
$Z_2$ orbifold symmetry if $\Psi_L(x,z)$ is an odd function of 
z and $\Psi_R(x,z)$ is even, or vice versa. To perform the 
Kaluza--Klein decomposition we write
\begin{equation}\label{KK}
   \Psi_{L,R}(x,z) = \sum_n \psi_n^{L,R}(x)\,
   e^{2\sigma}\,\hat f_n^{L,R}(z) \,.
\end{equation}
Because of the $Z_2$ symmetry of the action it is sufficient to 
restrict the integration over $z$ from $\phi$ to $\lambda$. Inserting the ansatz (\ref{KK}) 
into the action and requiring that the result takes the form of the 
usual Dirac action for massive fermions in four dimensions,
\begin{equation}\label{Sferm}
   S = \sum_n \int\!\mbox{d}^4x\,\sqrt{-g_4}\Big\{
   \bar\psi_n(x)\,i\rlap/\partial\,\psi_n(x)
   - m_n\,\bar\psi_n(x)\,\psi_n(x) \Big\} \,,
\end{equation}
where $\psi\equiv\psi_L+\psi_R$ (except for possible chiral modes) 
and $m_n\ge 0$ , we find that the functions $\hat f_n^{L,R}(\phi)$ 
must obey the conditions
\begin{eqnarray}\label{cond}
   \int\limits_\phi^\lambda\!\mbox{d}z\,e^\sigma
   \hat f_m^{L*}(z)\,\hat f_n^L(z)
   &=& \int\limits_\phi^\lambda\!\mbox{d}z\,e^\sigma
    \hat f_m^{R*}(z)\,\hat f_n^R(z) = \delta_{mn} \,,
    \nonumber\\
   \left( \pm\,\partial_z - m \right)
   \hat f_n^{L,R}(z) 
   &=& - m_n\,e^\sigma \hat f_n^{R,L}(z) \,.
\end{eqnarray}
The boundary conditions $\hat f_m^{L*}(\phi)\,\hat f_n^R(\phi)
=\hat f_m^{L*}(\lambda)\,\hat f_n^R(\lambda)=0$, which follow since either 
all left-handed or all right-handed functions are $Z_2$-odd, ensure
that the differential operators $(\pm \partial_z -m)$ 
are hermitian and their eigenvalues $m_n$ real. Since the equations 
are real, the functions $\hat f_n^{L,R}(\phi)$ could be chosen real 
without loss of generality.

\subsection{The right-handed zero mode: wave function localization}

We now choose boundary conditions such that all left-handed modes are odd under orbifold parity. 
We specialize our investigation to the right-handed massless zero-mode. (\ref{cond}) now takes the form
\begin{eqnarray}\label{cond2}
   \int\limits_\phi^\lambda\!\mbox{d}z\,e^\sigma
   \hat f_0^{R*}(z)\,\hat f_0^R(z)
   & = 1 \,,
    \nonumber\\
   \left( \partial_z + m \right)
   \hat f_0^{R}(z) 
   &=& 0 \,.
\end{eqnarray}

Let's consider, for example, the limit $\alpha\Rightarrow 0$, then $a(z)=e^{-kz}$ and we can write
\begin{eqnarray}
   \hat f^2(z)=\frac{k'e^{-2mz}}{e^{k'\lambda}-e^{k'\phi}},
\end{eqnarray}
where we defined $k'=k-2m$ and $\hat f=\hat f_0^R$.

The zero-mode wave function on the visible brane is thus given by
\begin{eqnarray}\label{massa}
   \hat f^2(\phi)=\frac{k'e^{-2m\phi}}{e^{k'\lambda}-e^{k'\phi}}.
\end{eqnarray}

Remarkably it is possible to localize the wave function on the hidden brane; in particular for well-separated brane we distinguish the cases:

\begin{itemize}
  \item $k'>0$ ................. $\hat f^2(\phi)<<1\Longleftrightarrow k'\lambda>>-2m\phi$
  \item $k'<0$ ................. $\hat f^2(\phi)<<1\Longleftrightarrow e^{-k\phi}<<1$.
  \end{itemize}

\subsection{The Higgs sector and the neutrino mass}
We now describe the realization of a small Dirac neutrino mass in four dimensions, using the localization of the wave 
function mentioned above. 
Omitting gauge interactions, the action for a Higgs doublet 
$H=(\phi_1,\phi_2)$, a left-handed lepton doublet $L=(\nu_L,e_L)$ 
and a right-handed lepton $e_R$ localized on the visible brane is 
\begin{eqnarray}
   S &=& \int\!\mbox{d}^4x\,\sqrt{-g^{\rm ind,1}} \left\{
    g_{\rm ind,1}^{\mu\nu}\,\partial_\mu H_B^\dagger\,\partial_\nu H_B
    - \lambda \left( |H_B|^2 - v_B^2 \right)^2 \right\} \nonumber\\
   &+& \int\!\mbox{d}^4x\,\sqrt{-g^{\rm ind,1}} \left\{
    \bar L_B\hat\gamma^\mu\partial_\mu L_B
    + \bar e_{RB}\hat\gamma^\mu\partial_\mu e_{RB}
    - \left( y_e \bar L_B H_B e_{RB} + \mbox{h.c.} \right)
    \right\} \,,
\end{eqnarray}
where $g_{\rm ind,1}^{\mu\nu}$ is the 
induced metric on the visible brane, and the superscript over the gamma 
matrices is used to make the vielbein implicit.

We now introduce a Yukawa coupling of the bulk fermion with the Higgs 
and lepton fields. With our choice of boundary conditions all 
left-handed Kaluza--Klein modes vanish at the visible brane, so only 
the right-handed modes can couple to the Standard Model fields on the 
brane. We consider a coupling of the form:
\begin{equation}
   S_Y = - \int\!\mbox{d}^4x\,\sqrt{-g_{\rm ind,1}} \left\{
   \hat Y_5 \bar L_B(x) \widetilde H_B(x) \Psi_R(x,\phi)
   + \mbox{h.c.} \right\} \,,
\end{equation}
where $\widetilde H=i\sigma_2 H^*$, and the Yukawa coupling $\hat Y_5$ 
is naturally of order $M^{-1/2}$, with $M$ the fundamental Planck 
scale of the theory. Rescaling the Standard Model fields and inserting for the bulk fermion the Kaluza--Klein 
ansatz (\ref{KK}), we find
\begin{equation}
   S_Y = - \sum_{n\ge 0} \int\!\mbox{d}^4x \sqrt{-g_{\rm ind,1}}\left\{ \frac{Y_5}{a^2(\phi)} 
   \bar L_B(x) \widetilde H_B(x) \psi_n^R(x) f_n^R(\phi) + \mbox{h.c.} \right\}
\end{equation}
where $Y_5$ is naturally of order unity.

Specializing this formula to the right-handed zero mode we have
\begin{equation}
   S_Y = - \int\!\mbox{d}^4x \sqrt{-g_{\rm ind,1}}\left\{ \frac{Y_5}{a^2(\phi)} 
   \bar L_B(x) \widetilde H_B(x) \psi_0^R(x) f_0^R(\phi) + \mbox{h.c.} \right\}
\end{equation}
with a mass for the 4d neutrino that can be expressed as a function of the moduli in the form
\begin{equation}
   m_{\nu} = m_{\nu}(\phi,\lambda)=\frac{Y_5f_0^R(\phi)v_b}{a^2(\phi)},
\end{equation}
where $f_0^R(\phi)$ is given by (\ref{massa}) after the rescaling.
\subsection{Conformal transformation}
Now we want to study the effect of a conformal transformation to the Einstein frame ($\tilde g$) on the neutrino mass. In order 
to do that, it is necessary to determine the rescaling laws of the various fields in the Yukawa term.
We have to take into account two different conformal transformations.
\begin{itemize}
 \item $g^{ind}\Longrightarrow  g^4$  
 \end{itemize}
The transformations is completely defined by equations (19) and (20).

\begin{itemize}
 \item $g^4\Longrightarrow \tilde g$ ..............  $ \tilde g_{\mu\nu} = Q^2 g_{\mu\nu}^4$
 \end{itemize}

Collecting the two transformations together and remembering $\beta = \frac{2\alpha^2 + 1}{4\alpha^2}$ we have:
\begin{eqnarray}
g_{\mu\nu}^{ind,1}=a^2(\phi)g_{\mu\nu}^4=a^2(\phi) Q^{-2} \tilde g_{\mu\nu} \equiv A^2 \tilde g_{\mu\nu}; \\
g_{\mu\nu}^{ind,2}=a^2(\lambda)g_{\mu\nu}^4=a^2(\lambda) Q^{-2} \tilde g_{\mu\nu} \equiv B^2 \tilde g_{\mu\nu};
\end{eqnarray}
where 

\begin{equation}\label{fattori}
A=Q^{-\frac{1}{2 \beta}}(\cosh R)^{\frac{1}{1+2 \alpha^2}},\ B=Q^{-\frac{1}{2 \beta}}(\sinh R)^{\frac{1}{1+2 \alpha^2}}.
\end{equation}

From these formulas, requiring a canonical normalization of the kinetic terms of the fields, we find the following conformal rescalings for the fields:
\begin{itemize}
 \item CHIRAL ZERO MODE: $\Psi_0^R\Longrightarrow \tilde \Psi_0^R=Q^{-3/2} \Psi_0^R$  
 \item HIGGS FIELD: $H_B\Longrightarrow H_E=\frac{a(\phi)}{Q}  H_B$
 \item STANDARD MODEL'S SPINOR FIELD: $\Psi\Longrightarrow \tilde \Psi=(\frac{a(\phi)}{Q})^{3/2} \Psi$.
 \end{itemize}

Using these last formulas in the Yukawa term, we obtain the "correction" to 
the neutrino mass induced by the conformal transformation to the Einstein frame:
\begin{equation}
\tilde m_{\nu}(\phi,\lambda)=\frac{a(\phi)^{5/2}}{Q} m_{\nu}(\phi,\lambda).
\end{equation}

\section{The moduli potential}

In the original two-branes model proposed by Randall and Sundrum (\cite{RS1},\cite{RS2}) the brane tensions were 
tuned in order to have Minkowski branes.
The separation between the branes becomes a field (radion) in the process of 
compactification to four dimensions. With flat branes the classical potential 
for the radion is absent: the interbrane distance is parameterized by a massless scalar field. Since 
this field has not been observed experimentally, the authors were faced with the so-called "radion stabilization problem". One 
possible solution is 
provided by quantum vacuum energy of bulk fields: Casimir contribution of bulk fields generates a potential for the radion. 
The possibility of radion stabilization by this mechanism has been considered by several authors using scalar 
fields (\cite{scalarf},\cite{GPT}, \cite{FT}), spinor fields (\cite{FMT,FMT2}). 
The result is that it is generally not possible to give an acceptable mass to the radion and solve the hierarchy problem simultaneously. 

When the branes are $dS_4$ or $AdS_4$ a classical potential for the radion is present. 
It is stabilizing for the $AdS_4$ case and unstable for $dS_4$ branes \cite{baggerredi,detuned}. 
The Casimir effect is then a quantum correction to the classical potential (\cite{baggerredi}). In the $dS_4$ 
brane case the Casimir contribution has been calculated for conformally 
coupled scalar fields \cite{WN}, for massless spinor fields (\cite{fermionc}) and for massive scalar fields (\cite{ENOO}). 
The $AdS_4$ case has been studied in \cite{norman}.

In this section we will present the classical potential for the moduli (Q, R) in the $dS_4$ case.
Then we proceed with the evaluation of the quantum correction induced by a massive spinor field in the bulk. 
\subsection{The classical potential}

In order to write the classical potential it is necessary to review the main characteristics of the
supersymmetric Randall-Sundrum model (\cite{susyrs}) with detuned brane tensions. The detuning process will render our model more
realistic: since we want to describe an Universe with a positive cosmological constant, a de Sitter geometry for branes is 
the correct one and detuning is necessary. 

Denoting $T_{1,2}$ and $T_f$ respectively
the tension of the first (second) brane and the fine-tuned tension, supersymmetry imposes the bound $|T_{1,2}| \le T_f$.
The detuning process converts Minkowski branes in curved branes. The $AdS_4$ solution arises when $|T_{1,2}| < T_f$; the $dS_4$
solution corresponds to $|T_{1,2}|> T_f$.

When one detunes the brane tension $U_B\to T U_B$, the moduli
pick up a potential \cite{brax, marsiglia}:

\begin{equation}
V = \frac {6(T-1)k}{\kappa_5^2} e^{\alpha \psi}.
\end{equation}
Taking into account both branes, the classical potential expressed as a function of $S=lnQ$ and $R$ in the Einstein frame is

\begin{equation}\label{poti}
V_{class}(S,R) = V_{eff} + W_{eff},
\end{equation}
where
\begin{equation}\label{poti}
V_{eff}(S,R) = \frac{6(T-1)k}{\kappa_5^2} e^{-12\alpha^2 S/(1+2\alpha^2)}
\left(\cosh R \right)^{(4-4\alpha^2)/(1+2\alpha^2)}
\end{equation}
and
\begin{equation}\label{poti2}
W_{eff}(S,R) = \frac{6(T-1)k}{\kappa_5^2} e^{-12\alpha^2 S/(1+2\alpha^2)}
\left(\sinh R \right)^{(4-4\alpha^2)/(1+2\alpha^2)}.
\end{equation}
Note that $T=1$ corresponds to the BPS case.

As mentioned in \cite{baggerredi} the potential is unstable for $dS_4$ and we proceed with the analysis of the 
Casimir energy as a stabilizing contribution for the system.
\subsection{Casimir contribution: the 5-dimensional bag}
We will calculate the Casimir contribution of the bulk spinor on a Euclideanised form of the metric. On the 
Euclidean section the de 
Sitter branes become concentric four spheres \cite{GS},
\begin{equation}
ds^2 =dz^2+a^2(z)d\Omega^2_4,
\end{equation}
where $d\Omega^2_4$ is the 4-sphere metric.
The metric is conformal to a cylinder $I\times S^4$ \cite{GS,GEN}. Thus,
\begin{equation}
ds^2 =a^2(z)(dy^2+d\Omega^2_4)
\quad\qquad a(z)=(1-4k\alpha^2 z)^{\frac{1}{4 \alpha^2}}\,,
\label{metric}
\end{equation}
where the coordinates are dimensionful and we defined $dy=\frac{dz}{a(z)}$.
The dimensional length $L$ is given by
\begin{equation}
L(\phi,\lambda)=\int_{\phi}^{\lambda}\frac{dz}
{a(z)}=-\frac{1}{k} \frac{1}{4 \alpha^2 -1} [(1-4k \alpha^2 \lambda)^{1-\frac{1}{4 \alpha^2}}-(1-4k \alpha^2 \phi)^{1-\frac{1}{4 \alpha^2}}].
\end{equation}

Let's consider a second transformation given by

\begin{equation}\label{trconf}
y=-R lnr,
\end{equation}
with $0<r<1$ and $R$ is a constant introduced for dimensional reasons. In this way (\ref{metric}) becomes

\begin{equation}
ds^2 =\frac{a^2(z)}{r^2}(dr'^2 + r'^2 d\Sigma^2) \equiv \beta^2 [dr'^2 + r'^2 d \Sigma^2],
\label{metric2}
\end{equation}
where $d \Sigma^2 \equiv R^{-2} d \Omega_4^2$ is the metric for the 4-sphere of radius one, $\beta^2 \equiv a^2/r^2$ is the total conformal
factor and $r' \equiv Rr$ is the dimensionful radial coordinate ($0<r'<R$). The metric (\ref{metric}) is thus conformally related
to a 5-dimensional generalized cone endowed with 4-sphere as a base: a 5-ball of radius R. 

In order to keep track of the geometrical "reconfiguration" of the system under the total conformal 
transormation, let's consider the following formulas:

\begin{itemize}
\item $z<z^* \equiv \frac{1}{4k \alpha^2}$ where $z^*$ corresponds to the position of the bulk singularity,
\item $y \equiv -\frac{1}{k} \frac{1}{4 \alpha^2 -1} (1-4k \alpha^2 z)^{1-\frac{1}{4 \alpha^2}}$,
\item $y=-Rlnr$.
\end{itemize}
The first one corresponds to the assumption that the branes can't fall into the bulk singularity.
Since we are interested in the description of a low-redshift Universe, we will consider well-separated branes:
we assume that the hidden brane is close to the bulk singularity ($z_{hidden} \sim z^*$, remarkably this is a general behaviour 
in the low-energy regime of 
BPS-braneworlds \cite{palma}) while the visible one 
is located far away in the bulk ($z_{vis}<<z_{hidden}$). 
The remaining formulas modify the branes coordinate 
in the following way:
\begin{eqnarray} 
z \rightarrow -\infty  \Longleftrightarrow y=0 \Longleftrightarrow r=1,  VISIBLE \\
z \rightarrow z^* \Longleftrightarrow y\rightarrow +\infty \Longleftrightarrow r=0, SINGULARITY.
\end{eqnarray}

Remarkably the total conformal transformation translated the cosmological two-branes
set up into a 5-ball. The ball radius is connected to the moduli $\lambda$ and $\phi$ by 

\begin{equation}
\frac{L(\lambda,\phi)}{R}=-ln \epsilon,
\end{equation}
where $\epsilon$ is a dimensionless parameter that is small for well-separated branes. In this geometrical configuration 
the evaluation of the effective action will be easier, as we now deduce.

We will start recalling the SPINOR action and studying the effect of the conformal transformation on 
the spinor field $\Psi$. The action is

\begin{equation}\label{action}
   S = \int\!\mbox{d}^4x\!\int\!\mbox{d}z\,\sqrt{-g_5}
   \left\{ E_a^A \left[ \frac{i}{2}\,\bar\Psi\gamma^a
   (\partial_A-\overleftarrow{\partial_A})\Psi
   + \frac{\omega_{bcA}}{8}\,\bar\Psi \{\gamma^a,\sigma^{bc}\} \Psi
   \right] - m\,\mbox{sgn}(\alpha)\,\bar\Psi\Psi \right\} \,.
\end{equation}

In the transformation $g_{AB}=\beta^2 g'_{AB}$ the action is rewritten as

\begin{equation}\label{apice}
   S = \int\!\mbox{d}^5x\,\sqrt{-g'_5}
   \left\{ E_a^{A'} \left[ \frac{i}{2}\,\bar\Psi'\gamma^a
   (\partial_A-\overleftarrow{\partial_A})\Psi'
   + \frac{\omega_{bcA}}{8}\,\bar\Psi' \{\gamma^a,\sigma^{bc}\} \Psi'
   \right] - m'\,\mbox{sgn}(\alpha)\,\bar\Psi'\Psi' \right\} \,.
\end{equation}
where the primed quantities are referred to the ball metric and they are given by
$\Psi'=\beta^{2} \Psi$, $m'=\beta m$ and $E_a^{A'}=\beta E_a^{A}$.
We will now use a more compact notation and, omitting the prime, we write (\ref{apice}) as
\begin{equation}
   S = i \int\!\mbox{d}^5x\ \Psi^* D \Psi
\end{equation}
where $D \equiv \slash{\!\nabla}+ im$.

To proceed further we need to specify boundary conditions for the bulk fermion. 
Since we are interested in a chiral theory on the branes, we choose the "option-L" of \cite{grossman}, namely: all the left handed modes are 
odd under orbifold parity. In this way the correct boundary condition is 

\begin{eqnarray} 
z=\phi :\quad P_-\psi=0,
z=\lambda :\quad P_-\psi=0,
\end{eqnarray}
where $P_{\pm}=\frac12(1 \pm \gamma_5)$. To complete the specification of the 
boundary conditions we remember the existence condition for the operator $D^*$: if $P_-\psi=0$, 
this requires that the normal derivative $(\partial_y - m)P_+ \Psi=0$
should vanish. In summary, we have defined two subspaces in direct sum generated by 
the operators $P_{\pm}$; while on the first space (-) we have Dirichlet condition, on the second one (+) 
we impose Robin boundary condition. In other words the bulk fermion satisfies mixed boundary conditions (for 
further details see \cite{FMT,Mth,MOSS}).

In this way our cosmological set up is the 5-dimensional "extension" of the MIT bag model \cite{mit}. In these 
systems quarks and gluons
are free inside the bag, but they are unable to cross the boundary. This last condition 
corresponds precisely to the mixed boundary
conditions discussed above: a chiral theory on the branes corresponds to the condition 
that no quark current is lost through the boundary. The zeta function for massive fermionic fields inside
the bag has been considered in \cite{bag}. Analogous calculations have been developed for massive scalar field (\cite{bk,bek,bekl96}).
Functional determinants were discussed in \cite{begk,bdk,abdk,eli1,eli2,stuart,Mth}. 
For moduli stabilization with zeta function regularization see also \cite{GPT2,FGPT}. 

In the next section we will discuss the zeta function of the system following the approach of \cite{bag}.

\subsection{Casimir contribution: the $\zeta$ function}

We start recalling the general set-up for the MIT bag model in three dimensions. The setting we 
consider first is the Dirac spinor inside a spherically symmetric bag
confined to it by the appropriate boundary conditions. 
 Thus, we must solve the equation:
\beq
H\phi_n (\vec r) = E_n \phi_n (\vec r) , \label{2.1}
\eeq
$H$ being the Hamiltonian, with the boundary conditions
\beq
\left[1+i \left(\frac{\vec r } r \vec \gamma \right) \right] \phi_n 
\left|_{r=R} \right. =0 .\label{2.3}
\eeq

The  separation to be carried out
 in the eigenvalue equation (\ref{2.1}) is rather
standard and will not be given here in detail. Let $\vec J$ be the 
total angular momentum operator and $K$ the spin projection operator. 
Then there exists a simultaneous set of eigenvectors of $H,\vec J^2,J_3,K$
and the parity $P$. The eigenfunctions for positive eigenvalues $\kappa
=j+1/2$ of $K$ read
\beq
\phi_{jm} = \frac A {\sqrt{r} } \left(
   \begin{array}{l} i J_{j+1} (\omega r ) \Omega_{jlm} \left( \frac {\vec r} r 
\right)  \\
    - \sqrt{\frac{E-m}{E+m} } J_j (\omega r ) \Omega_{jl'm } 
         \left( \frac {\vec r} r \right)
   \end{array} 
\right), \label{2.4}
\eeq
whereas, for $\kappa = -(j+1/2) $, one finds
\beq
\phi_{jm} = \frac A {\sqrt{r} } \left(
   \begin{array}{l} i J_{j} (\omega r ) \Omega_{jlm} \left( \frac {\vec r} r
\right)  \\
     \sqrt{\frac{E-m}{E+m} } J_{j+1} (\omega r ) \Omega_{jl'm }
         \left( \frac {\vec r} r \right)
   \end{array}
\right). \label{2.5}
\eeq
Here 
$\omega=\sqrt{E^2-m^2}$, $A$ is a normalization constant and 
$\Omega_{jlm} (\vec r /r)$ are the well known spinor harmonics. 
In order to obtain eigenfunctions of the parity operator we must set
 $l'=l-1$ in
(\ref{2.4}) and $l'=l+1$ in (\ref{2.5}). In both cases, $j=1/2,3/2,...,
\infty$, and the eigenvalues are degenerate in $m=-j,...,+j$.

Imposing the boundary conditions (\ref{2.3}) on the solutions 
(\ref{2.4}) and (\ref{2.5}), respectively, one easily finds the 
corresponding implicit eigenvalue
equation. For $\kappa > 0$, it reads
\beq
\sqrt{\frac{E+m}{E-m}} J_{j+1} (\omega R) +J_j (\omega R) =0 , \label{2.6}
\eeq
and for $\kappa < 0$, on its turn, 
\beq
J_j (\omega R)- \sqrt{\frac{E-m}{E+m}} J_{j+1} (\omega R) =0 .\label{2.7}
\eeq

We define the zeta function of the system as
\beq
\zeta (s) = \sum_k (E_k^2) ^{-s} \label{relzeta}
. 
\eeq
The power of the method lies in the well defined prescriptions 
and procedures that we have at our hand to analytically continue the series
to the rest of the complex $s$-plane, even when the spectrum $E_k$ is
not known explicitly. In particular we are interested in the calculation of the Casimir energy for
massive spinors in the bag and we will employ zeta function to regularize this ground state energy. Briefly we consider
\beq\label{introcas}
E_0 (s) &=& - \frac{1}{2} \sum_k 
\left( E_k^2 \right)^{1/2 -s} \mu^{2s}, \qquad
\mbox{Re}\ s >s_0= 2\nonumber\\
&=& -\frac 1 2 \zeta (s-1/2) \mu^{2s} \label{grounden},
\eeq
and later analytically continue to the value $s=0$ in the complex plane.
Here $s_0$ is the abscissa of convergence of the series, $\mu$ the usual 
mass parameter.

Let's apply this QCD analysis to the cosmological branes system. When we pass from three to five dimensions
we have to take into account new degeneracy factors for the eigenvalues, while the implicit eigenvalue equation 
remains formally untouched. In particular the degeneracy $d(j)$ for a spinor field on the ball is
\begin{equation}
d(j)=d_s {n+D-2\choose D-2},
\end{equation}
where $d_s$ is the spinor dimension, $D$ is the manifold dimension (i.e. the dimension of the generalized cone), $j=n+ \frac{D}{2} -1$ and $n=0,1,2,...$.
Since a closed analytical form for the eigenvalues is not available for this system, we will 
exploit the residue theorem (for a pedagogical description of these techniques see \cite{kirsten}):
\beq
\zeta (s) &=&  \sum_{j=3/2,\ldots}^\infty d(j) \int_\gamma \frac{dk}{2\pi i} 
(k^2 +m^2)^{-s}  \nn \\ && \hspace{-10mm} 
\times \, \frac{\partial}{\partial k}
 \ln \left[ J_j^2 (kR) - J_{j+1}^2 (kR) 
+ \frac{2m}{k} J_j (kR) J_{j+1} (kR) \right].
\eeq

Using the method ---ordinarily employed in this situation--- of deforming the 
contour which originally encloses the singular points on the real axis,
until it covers the imaginary axis, 
after some manipulations we obtain the following equivalent expression for 
$\zeta$:
\beq
\zeta   (s) &=& \frac{ \sin \pi s}{\pi} 
\sum_{j=3/2,\ldots}^\infty d(j) 
\int_{mR/j}^\infty dz \, \left[ \left( \frac{zj}{R}\right)^2 -m^2
\right]^{-s} \nn  \\
&& \times \frac{\partial}{\partial z} \ln\left\{z^{-2j} \left[ I_j^2 (zj)
 \left( 1 + \frac{1}{z^2}- \frac{2mR}{z^2j} \right) + {I_j'}^2 (zj)
\right. \right. \nn \\ && \left. \left.
 + \frac{2R}{zj} \left( m - \frac{j}{R} \right) I_j (zj) {I_j}' (zj)
\right]\right\},
\eeq
where we defined $z=k/j$.
As is usual, we will now split the zeta function into two parts:
\beq
\zeta (s)=Z_N(s)+\sum_{i=-1}^N A_i(s),
\eeq 
namely a regular one, $Z_N$, and a remainder that contains the 
contributions of the $N$ first terms
of the Bessel functions $I_\nu (k)$ as $\nu, k \rightarrow \infty$ with 
$\nu /k$ fixed.
The regular part of the zeta function is:
\beq
Z_N(s)&=& \sn \smj \itz \paran^{-s}  \nn\\
&& \hspace{-17mm}\times \frac{\partial}{\partial z}
 \left\{ \ln \left[I_j^2(zj)(1+\frac{1}{z^2}-\frac{2mR}{z^2j})+
I'_j\!{}^2(zj)+\frac{2R}{zj} (m-\frac{j}{R}) I_j(zj)I'_j(zj)\right] \right. 
 \nn\\
&&-\left.\ln\left[\frac{e^{2j\eta}(1+z^2)^\frac{1}{2}(1-t)}{\pi j z^2}\right]
   -\sum_{k=1}^N \frac{D_k(mR,t)}{j^k}\right\},
\eeq

where $\eta = \sqrt{1+z^2} +\ln [z/(1+\sqrt{1+z^2})]$ and $t=1/\sqrt{1+z^2}$.
After renaming $mR=x$, the relevant polynomials (see appendix A) are given by
\beq
D_1(t)&=&{\frac {{t}^{3}}{12}}+\left (x-1/4\right )t \nn\\
D_2(t)&=&-{\frac {{t}^{6}}{8}}-{\frac {{t}^{5}}{8}}+\left (-{\frac {x}{2}}+1/8
\right ){t}^{4}+\left (-{\frac {x}{2}}+1/8\right ){t}^{3}-{\frac {{t}^
{2}{x}^{2}}{2}} \nn\\
D_3(t)&=&{\frac {179\,{t}^{9}}{576}}+{\frac {3\,{t}^{8}}{8}}+\left (-{\frac {23
}{64}}+{\frac {7\,x}{8}}\right ){t}^{7}+\left (x-1/2\right ){t}^{6}+
\left ({\frac {9}{320}}-{\frac {x}{4}}+{\frac {{x}^{2}}{2}}\right ){t}
^{5}    \nn\\
&&+\left ({\frac {{x}^{2}}{2}}+1/8-{\frac {x}{2}}\right ){t}^{4}+
\left (-{\frac {x}{8}}+{\frac {5}{192}}+{\frac {{x}^{3}}{3}}\right ){t
}^{3} \nn\\
D_4(t)&=& -\frac{71}{64} t^{12} - \frac{179}{128} t^{11} + (\frac{57}{32} - \frac{21}{8} x)t^{10} + (\frac{327}{128} - \frac{49}{16} x) t^{9} 
+ (-\frac{37}{64} + 2x - x^2) t^{8}    \nn\\
&&+ (-\frac{165}{128} + 3x - \frac{5 x^2}{4})t^7 + (-\frac{1}{8} + \frac{x}{4} +\frac{x^2}{8} -\frac{x^3}{2})t^6 + (\frac{17}{128} - \frac{7x}{16} 
+ \frac{x^2}{2}- \frac{x^3}{2})t^5    \nn\\
&&+ (\frac{1}{32} - \frac{x}{8} + \frac{x^2}{8} - \frac{x^4}{4})t^4  \nn\\
D_5(t)&=& \frac{40573}{7680} t^{15} + \frac{213}{32} t^{14} + ( -\frac{5853}{512} + \frac{1463x}{128})t^{13} + (\frac{105x}{8}-\frac{507}{32})t^{12}  \nn\\
&& + (\frac{1835}{256}- \frac{487x}{32} + \frac{49 x^2}{16}) t^{11} + (\frac{397}{32}-\frac{159x}{8}+ 4 x^2 )t^{10} + (-\frac{1441}{2304}+\frac{217x}{64}-
\frac{13x^2}{8} + \frac{9 x^3}{8}) t^{9} \nn\\
&& + (-\frac{109}{32} + \frac{31x}{4} - \frac{27x^2}{8} + \frac{3x^3}{2}) t^{8} + 
(-\frac{1567}{3584} + \frac{33x}{32} - \frac{9x^2}{16} + \frac{x^4}{2}) t^{7}  \nn\\
&& + (-\frac{x}{2} + \frac{3x^2}{8} - \frac{x^3}{2} + \frac{3}{16} +  \frac{x^4}{2}) t^{6} + (\frac{107}{2560} - \frac{17x}{128} + \frac{x^2}{8} - \frac{x^3}{8} + \frac{x^5}{5})t^5.
\label{asympol}
\eeq
The number N of terms to be subtracted must be high enough in order to absorb all possible divergent
contributions into the groundstate energy. In our case N=5.
The asymptotic contributions $A_i (s)$, $i=-1,...,5$ are defined as
\beq
A_{-1} (s) &=& 
\frac{\sin (\pi s)} {\pi} \sum_{j=3/2}^{\infty} 
2jd(j) \int_{mR/j} ^{\infty} dz \left( \left( \frac{zj} R \right)^2 -m^2 
\right)^{-s}
\frac{\sqrt{1+z^2} -1} z \nonumber\\
A_0 (s) &=& 
\frac{\sin (\pi s)} {\pi} \sum_{j=3/2}^{\infty}
d(j) \int_{mR/j} ^{\infty} dz \left( \left( \frac{zj} R \right)^2 -m^2 
\right)^{-s}
\frac{\partial} {\partial z} \ln \frac{\sqrt{1+z^2} (1-t) } {z^2} 
\nn\\
A_i (s) &=& 
\frac{\sin (\pi s)} {\pi} \sum_{j=3/2}^{\infty}
d(j) \int_{mR/j} ^{\infty} dz \left( \left( \frac{zj} R \right)^2 -m^2 
\right)^{-s}
\frac{\partial} {\partial z}
\frac{D_i (t)} {j^i}.  \label{aisanf}
\eeq
This completes the description of the zeta function. In the next section we will consider the divergent contributions
in the vacuum energy. After that we will move to the renormalization process.
\subsection{Casimir contribution: Residue [$\zeta(-1/2)$]}

As a first step towards renormalization, let's start considering the divergent terms in the groundstate energy.
By construction they are contained in the functions $A_i(s)$. We will discuss them separately following the approach 
of \cite{bag},\cite{bekl96}.

\subsubsection{Residue $A_{-1}(-1/2)$}

The first asymptotic contribution reads
\beq
A_{-1} (s) &=& 
\frac{\sin (\pi s)} {\pi} \sum_{j=3/2}^{\infty} 
2jd(j) \int_{mR/j} ^{\infty} dz \left( \left( \frac{zj} R \right)^2 -m^2 
\right)^{-s}
\frac{\sqrt{1+z^2} -1} z
\eeq

of which we need the analytical continuation to $s=-1/2$.
With the substitution $t=(zj /R)^2 -m^2$, this expression results into
the following one
\begin{eqnarray}
A_{-1} (s)
&=& \frac{\sin (\pi s)}{\pi} \sum_{j =3/2}^{\infty} d(j) \int\limits _0 ^{\infty}
dt\,\,\frac{t^{-s}}{t+m^2}\left\{
\sqrt{j^2 +R^2 (t+m^2)} -j\right\}\nonumber\\
&=&-\frac 1 {2\sqrt{\pi}}
 \frac{\sin (\pi s)}{\pi} \sum_{j=3/2}^{\infty} d(j) \int\limits _0 ^{\infty}
dt\,\,t^{-s} \int\limits _0^\infty
d\alpha \,\, e^{-\alpha (t+m^2)}\label{anhdaa2}\\
& &\times\int\limits_0^{\infty}d\beta \,\,\beta ^{-3/2} \left\{
e^{-\beta (j^2 +R^2 [t+m^2] ) } -e^{-\beta j^2}\right\},\nonumber
\end{eqnarray}
where the Mellin integral representation for the single factors has been
used. As we see,
the $\beta$-integral is well defined. In order to deal separately 
with the two contributions in the $\beta$-integral, we introduce a regularization parameter
$\delta$. In this way $A_{-1}(s)$ can be written as
\begin{eqnarray}
A_{-1} (s) =\lim_{\delta \to 0} \left[ A_{-1}^1 (s,\delta )
+A_{-1} ^2 (s,\delta ) \right], \label{anhdaa3}
\end{eqnarray}
with
\begin{eqnarray}
A_{-1}^1 (s,\delta )
 &=&-\frac 1 {2\sqrt{\pi}}
 \frac{\sin (\pi s)}{\pi} \sum_{j=3/2}^{\infty} d(j)
\int\limits_0^\infty  d\alpha \,\,
 e^{-\alpha m^2}
\int\limits_0^{\infty}d\beta \,\,\beta ^{-3/2+\delta}e^{-\beta (j^2 +R^2
m^2 ) }
\int\limits _0 ^{\infty}
dt\,\,t^{-s}e^{-t(\alpha +\beta R^2)}\nonumber
\end{eqnarray}
and
\begin{eqnarray}
A_{-1} ^2 (s,\delta ) =    \frac 1 {2\sqrt{\pi}} \Gamma (1-s)
\frac{\sin (\pi s)}{\pi}    \sum_{j=3/2}^{\infty} d(j)
\int\limits_0^\infty  d\alpha \,\,
e^{-\alpha m^2 }   \alpha^{s-1}\int\limits_0^{\infty}d\beta \,\,
\beta ^{-3/2+\delta}e^{-\beta j^2}.
\nonumber
\end{eqnarray}

In $A_{-1}^1 (s,\delta
)$ two of the integrals can be done, yielding
\begin{eqnarray}
A_{-1}^1 (s,\delta )
&=& -\frac{R^{1-2\delta}}{2\sqrt{\pi}\Gamma (s)} \Gamma (s+\delta -1/2) \sum_{j=3/2}^{\infty} d(j) \int\limits  _1 ^{\infty}
dy\,\, y^{s -1} \left[ m^2 y + \left(\frac{j} R\right)^2
\right]^{1/2-s-\delta}. \label{anhdaa4}
\end{eqnarray}

Exploiting the identity

\beq
\int\limits_0^\infty  dx \,\,
\frac{x^{\alpha -1}}{(y+zx)^{s+ \epsilon}} = \frac{ \Gamma(\alpha) 
\Gamma( s + \epsilon - \alpha)}{\Gamma(s+ \epsilon)} z^{- \alpha} y^{- \epsilon -s + \alpha},
\eeq
it is possible to rewrite $A_{-1} ^2 (s,\delta )$ in the form
\begin{eqnarray}
A_{-1} ^2 (s,\delta )&=& \frac{R^{1-2\delta}}{2\sqrt{\pi}\Gamma (s)}
\Gamma (s+\delta -1/2)
\sum_{j=3/2}^{\infty} d(j) \int\limits  _0 ^{\infty}
dx\,\, x^{s-1        }
\left[ m^2 x +
\left(\frac{j} R\right)^2 \right]^{1/2-s-\delta}.
\label{anhdaa5}
\end{eqnarray}

Adding up (\ref{anhdaa4}) and (\ref{anhdaa5}) yields
\begin{eqnarray}
A_{-1} (s) &=& \frac R {2\sqrt{\pi}\Gamma (s)}
\Gamma (s -1/2) \sum_{j=3/2}^{\infty} d(j) \int\limits  _0 ^{1}
dx\,\, x^{s-1        }
\left[ m^2 x +
\left(\frac{j} R\right)^2 \right]^{1/2-s} \nn\\
&=& \frac{R^{2s}}{2\sqrt{\pi}} \frac{\G (s-1/2)}{\G (s+1)}
      \sum_{j=3/2}^\infty d(j) \left\{ \frac 1
     {\left[j^2 +(mR)^2\right] ^{s-1/2}} \right.\nn\\
& &\left.\hspace{1cm}\quad+\left( s-\frac 1 2 \right) (mR)^2
           \int_0^1 dx \,\, \frac{x^s}{(j^2 +(mR)^2 x)^{s+1/2}} \right\},
\label{anhdaa6}
\eeq
where in the last step a partial integration has been performed such that
the $x$-integral is well behaved at $s=-1/2$.

This is a form suited for the treatment of the sum,
which is performed using

\beq
\sum_{\nu =1/2,3/2,...}^{\infty} \nu^{2n+1} \left(1+\left(\frac{\nu} x
\right) ^2 \right) ^{-s} &=& \frac 1 2 \frac {n! \Gamma (s-n-1)}{\Gamma (s)}
x^{2n+2} \label{dispari}\\
& &\hspace{-25mm} +(-1)^n 2 \int\limits_0^x d\nu \,\, \frac{\nu^{2n+1} }
{1+e^{2\pi\nu}  } \left( 1-\left( \frac{\nu} x \right)^2 \right) ^{-s} \nn\\
& & \hspace{-25mm} +(-1)^n 2\cos (\pi s)
 \int\limits_x^{\infty} d\nu \,\, \frac{\nu^{2n+1} }
{1+e^{2\pi\nu}  } \left( \left( \frac{\nu} x \right)^2 -1
\right) ^{-s}\nn
\eeq
for odd powers of $\nu$ and

\beq \sum_{\nu =1/2,3/2,...}^{\infty}
\nu^{2n} \left(1+\left(\frac{\nu} x \right) ^2 \right) ^{-s} &=&
\frac 1 2 \frac { \Gamma (n+1/2) \Gamma (s-n-1/2)}{\Gamma (s)}
x^{2n+1} \label{pari}\\ & &\hspace{-25mm} -(-1)^n 2\sin (\pi s)
 \int\limits_x^{\infty} d\nu \,\, \frac{\nu^{2n} }
{1+e^{2\pi\nu}  } \left( \left( \frac{\nu} x \right)^2 -1
\right) ^{-s} \nn
\eeq
for even powers of $\nu$.
This formula as well as (\ref{dispari}) can only be applied for $Re(s)<1$, otherwise the integral diverges at the integration
limit $\nu=x$.

Since the prefactor in (\ref{anhdaa6}) contains a pole at s=-1/2 an expansion of the integrand is necessary and 
remembering that $d(j)=(2/3) (j^3 + \frac{3}{2} j^2 - \frac{j}{4} - \frac{3}{8})$, our final expression for the residue is found to be
\beq
\res A_{-1} (-1/2) &=& - \frac{457}{241920 \pi R} - \frac{17 m^2 R}{2880 \pi} + \frac{m^4 R^3}{144 \pi} + \frac{m^6 R^5}{180 \pi}.   
\eeq

\subsubsection{Residue $A_{i}(-1/2)$, $i\geq 1$}

Let's start remembering the form of the asymptotic contributions for $i\geq 1$:
\beq
A_i (s) &=& 
\frac{\sin (\pi s)} {\pi} \sum_{j=3/2}^{\infty}
d(j) \int_{mR/j} ^{\infty} dz \left( \left( \frac{zj} R \right)^2 -m^2 
\right)^{-s}
\frac{\partial} {\partial z}
\frac{D_i (t)} {j^i}.  
\eeq
We need to introduce some notation as follows:
 \beq 
D_i (t) =\sum_{a=0}^{2i}
x_{i,a} t^{i+a}
\label{eq2.26} 
\eeq
(where the relevant coefficients $x_{i,a}$ are explicitly listed in (\ref{asympol}) and 
\beq 
f(s;c;b) =\sum_{\nu =
1/2,3/2,\ldots} ^{\infty} j ^c \left( 1+ \left( \frac{j} {mR}
\right ) ^2 \right) ^{-s-b}.\nn 
\eeq
With this notation we write
\beq
A_i(s)&=& \frac{\sin (\pi s)} {\pi} \sum_{j=3/2}^{\infty}
\frac{d(j)}{j^i} \sum_{a=0}^{2i} x_{i,a} \int_{mR/j} ^{\infty} dz \left( \left( \frac{zj} R \right)^2 -m^2 
\right)^{-s}
\frac{\partial} {\partial z}
t^{a+i}=  \nn \\
&=& -\frac{\sin (\pi s)} {\pi} \sum_{a=0}^{2i} \left\{\frac{x_{i,a}m^{-2s}}{(mR)^{a+i}} \frac{\Gamma (s+ \frac{a+i}{2}) \Gamma (1-s)}{\Gamma (\frac{a+i}{2})} \sum_{j=3/2}^{\infty} d(j) j^a [1+(\frac{j}{mR})^2]^{-s-\frac{a+i}{2}}\right\}= \nn \\
&=& -\frac{\sin (\pi s) \Gamma (1-s)} {m^{2s} \pi} \frac{2}{3} \sum_{a=0}^{2i} \left\{
 \frac{x_{i,a}}{(mR)^{a+i}} \frac{\Gamma (s+ \frac{a+i}{2})}{\Gamma (\frac{a+i}{2})} 
\left[f(s;3+a;\frac{i+a}{2}) + \frac{3}{2} f(s;2+a;\frac{i+a}{2}) \right. \right. \nn \\
& & \left. \left. -\frac{1}{4} f(s;1+a;\frac{i+a}{2})-\frac{3}{8} f(s;a;\frac{i+a}{2}) \right]\right\},
\eeq
where we performed the z-integration using the identity (\cite{kirsten})
\beq 
\iinma t^n &=& -\mzs
\frac{n}{2(mR)^n}\frac{\g s+\frac n 2 \right) \Gamma (1-s)}{\g
1+\frac n 2\right)}  \nu ^n \left[1+\numr
\right]^{-s-\frac n 2}
\label{eq2.27} 
\eeq
and we exploited the vanishing of the degeneracy factor for $j=1/2$.
In order to evaluate the $f(s;a;b)$ for the relevant values in $s=-1/2$ we can use the following formulas:
\begin{itemize}
 \item the recurrence formula
 \beq
f(s;a;b) = (mR)^2 \left[ f(s;a-2;b-1)- f(s;a-2;b)\right];
\label{recur}
\eeq
 \item formulas \ref{dispari} and \ref{pari}, valid for $\Re{s}<1$;
 \item an extension of formulas \ref{dispari} and \ref{pari} to $\Re{s}<k+1\leq n+2$ obtained using partial integrations:
 \beq
 \sum_{\nu = 1/2}^{\infty} \nu^{2n+1} \left( 1+(\frac{\nu}{x})^2 \right) ^{-s} = 
\frac{n! \Gamma (s-n-1) x^{2n+2}}{2 \Gamma (s)} - \delta_{k,n+1} 
\frac{n! \Gamma (s-n-1)}{2 \Gamma (s)} x^{2n+2} \nn \\
+ 2(-1)^{k+n} \frac{\Gamma (s-k)}{\Gamma (s)} \int\limits_0^{x} d\nu \,\, \left[ \left( \frac{d}{d\nu} \frac{x^2}{2 \nu} \right)^k  \frac{\nu^{2n+1} }
{1+e^{2\pi\nu}  } \right] \left( 1- \left(  \frac{\nu} x \right)^2 
\right) ^{-s +k} \nn \\
+2(-1)^n cos(\pi s) \frac{\Gamma(s-k)}{\Gamma (s)} \int\limits_x^{\infty} d\nu \,\, \left[ \left( \frac{d}{d\nu} \frac{x^2}{2 \nu} \right)^k  \frac{\nu^{2n+1} }
{1+e^{2\pi\nu}  } \right] \left( \left( \frac{\nu} x \right)^2 -1 
\right) ^{-s +k}
 \eeq
 for odd powers of $\nu$ and
 \beq
  \sum_{\nu = 1/2}^{\infty} \nu^{2n} \left( 1+(\frac{\nu}{x})^2 \right) ^{-s} = 
\frac{\Gamma (n+1/2) \Gamma (s-n-1/2)}{2 \Gamma (s)} x^{2n +1} \nn \\
-2(-1)^n sin(\pi s) \frac{\Gamma(s-k)}{\Gamma (s)} \int\limits_x^{\infty} d\nu \,\, \left[ \left( \frac{d}{d\nu} \frac{x^2}{2 \nu} \right)^k  \frac{\nu^{2n} }
{1+e^{2\pi\nu}  } \right] \left( \left( \frac{\nu} x \right)^2 -1 
\right) ^{-s +k}
 \eeq
 for even powers of $\nu$.
\end{itemize}

In this way it is possible to construct an explicit form for the asymptotic contributions and, in particular,
our result for the residues is given by

\beq
\res A_{1} (-1/2) &=&  - \frac{17m}{2880 \pi} + \frac{17}{11520 \pi R} - \frac{m^2 R}{288 \pi} + \frac{m^3 R^2}{24 \pi} + \frac{m^4 R^3}{144 \pi} + \frac{m^5 R^4}{12 \pi}\nn\\
\res A_2 (-1/2) &=&  m (\frac{1}{64} + \frac{1}{16 \pi}) - \frac{1}{R} (\frac{1}{1024} + \frac{1}{192 \pi}) \nn\\
& &+ m^2 R (\frac{19}{512} + \frac{1}{48 \pi}) + 
(\frac{3}{32} + \frac{1}{4 \pi})m^3 R^2 + \frac{m^4 R^3}{16} \nn\\ 
\res A_{3} (-1/2) &=& -m (\frac{1}{192} + \frac{7}{480 \pi}) + \frac{1}{R} (\frac{1}{3072} + 
\frac{97}{241920 \pi}) -m^2 R (\frac{25}{1536} + \frac{565}{12096 \pi}) \nn \\
& & -m^3 R^2 (\frac{3}{32} + \frac{41}{120 \pi}) - m^4 R^3 (\frac{1}{16}+ \frac{2}{9 \pi}) - \frac{m^5 R^4}{9 \pi} \nn\\ 
\res A_{4} (-1/2) &=& -m (\frac{63}{4096} + \frac{11}{240 \pi}) + \frac{1}{R} (\frac{35}{65536} + 
\frac{13}{40320 \pi}) -m^2 R (\frac{13}{256} + \frac{1}{6 \pi}) \nn \\
& & -m^3 R^2 (\frac{3}{64} + \frac{1}{6 \pi}) - \frac{1}{32} m^4 R^3 \nn \\
\res A_{5} (-1/2) &=& m (\frac{61}{12288} + \frac{37}{2240 \pi})  + \frac{1}{R} (\frac{23}{196608}- \frac{31}{2661120 \pi}) + m^2 R (\frac{23}{768}+ \frac{359}{3780 \pi})\nn \\
& & + m^3 R^2 (\frac{3}{64} + \frac{181}{1260 \pi})  + m^4 R^3 (\frac{1}{32} + \frac{4}{45 \pi}) + \frac{2}{45 \pi} m^5 R^4.
\eeq

\subsubsection{Residue $A_{0}(-1/2)$}

This calculation will exploit the basic formulas we showed above and the Mellin integral representation.
With the substitution $t=(x \nu/a)^2 - m^2$ we write
\beq
A_0 (s) &=& 
\frac{\sin (\pi s)} {\pi} \sum_{j=3/2}^{\infty}
d(j) \int_{mR/j} ^{\infty} dz \left( \left( \frac{zj} R \right)^2 -m^2 
\right)^{-s}
\frac{\partial} {\partial z} \ln \frac{\sqrt{1+z^2} -1 } {z^2} = \nn \\
&=& - \frac{1}{2} \frac{\sin (\pi s)}{\pi} \sum_{j =3/2}^{\infty} d(j) \int\limits _0 ^{\infty}
dt\,\,\frac{t^{-s}}{t+m^2}\left\{\frac{
\sqrt{j^2 +R^2 (t+m^2)} -j}{\sqrt{j^2 +R^2 (t+m^2)}}\right\} \nn \\
&=& \frac 1 {2\sqrt{\pi}}
 \frac{\sin (\pi s)}{\pi} \sum_{j=3/2}^{\infty} d(j)j \int\limits _0 ^{\infty}
dt\,\,t^{-s} \int\limits _0^\infty
d\alpha \,\, e^{-\alpha (t+m^2)}\label{anhdaa2}\\
& &\times\int\limits_0^{\infty}d\beta \,\,\beta ^{-1/2} \left\{
e^{-\beta (j^2 +R^2 [t+m^2] ) } -e^{-\beta j^2}\right\},\nonumber
\eeq
where in the last step the Mellin representation for the single factors has been used
and, in particular, we exploited 
\beq
\frac{
\sqrt{j^2 +R^2 (t+m^2)} -j}{\sqrt{j^2 +R^2 (t+m^2)}}  = -\frac{j}{\sqrt{\pi}} \int\limits_0^{\infty}d\beta \,\,\beta ^{-1/2} \left\{
e^{-\beta (j^2 +R^2 [t+m^2] ) } -e^{-\beta j^2}\right\}.
\eeq

In this way we can closely follow the calculation discussed previously for $\mbox{Res}\ A_{-1} (-1/2)$.
The asymptotic contribution is rewritten as
\beq
A_{0} (s) &=&- \frac{R^{2s}}{2\sqrt{\pi}} \frac{\G (s+1/2)}{\G (s+1)}
      \sum_{j=3/2}^\infty d(j)j \left\{ \frac 1
     {\left[j^2 +(mR)^2\right] ^{s+1/2}} \right.\nn\\
& &\left.\hspace{1cm}\quad+\left( s+\frac 1 2 \right) (mR)^2
           \int_0^1 dx \,\, \frac{x^s}{(j^2 +(mR)^2 x)^{s+3/2}} \right\}
\label{azero}
\eeq
and the relevant residue is given by
\beq
\mbox{Res}\ A_0 (-1/2) = \frac{17}{1920 \pi R} + \frac{m^2 R}{16 \pi} + \frac{ m^4 R^3}{24 \pi}.
\label{poles}
\eeq

\subsection{Casimir contribution: renormalization}
For the discussion of the renormalization let us look for the divergent
terms in the groundstate energy. Adding up the contributions calculated previously, the residue 
for the zeta function is
\beq
\mbox{Res}\ \zeta (-1/2) = \frac{4m}{315 \pi} + \frac{41}{10395 \pi R}  - \frac{2 m^2 R}{45 \pi} 
-\frac{23 m^3 R^2}{315 \pi} - \frac{7 m^4 R^3}{90 \pi} + \frac{m^5 R^4}{60 \pi} + \frac{m^6 R^5}{180 \pi}
\label{poles}
\eeq
These terms form the minimal set of counterterms necessary in order to
renormalize our theory.
 
We are led into a physical system consisting of two parts:
\begin{enumerate}
\item A classical system consisting of a spherical surface ('bag') with radius
 $R$. Its energy reads:
\beq
E_{class} = a R^5 + b R^4 + c R^3 + d R^2 + e R + f + \frac{g}{R}.
\label{n1}
\eeq
This energy is determined by the parameters: $a, b, c, d, e, f, g$.
\item A spinor quantum field $\Psi (x)$ obeying the Dirac 
equation 
and the MIT boundary conditions 
(\ref{2.3}) on the surface.  
The quantum field has a ground state energy given by $E_0$, Eq.
(\ref{grounden}).    
\end{enumerate}
Thus, the complete energy of the physical system is
\beq
E= E_{class} +E_0 \label{eges}
\eeq
and in this context the renormalization can be achieved by shifting the 
parameters in $E_{class}$ by an amount which
cancels the divergent contributions. 

First we perform a kind of minimal subtraction, where only the 
divergent contribution is eliminated,
\beq
a &\to &a + \frac{m^6} {360 \pi}
\frac 1 {s}
\quad
b  \,\to  \, b +\frac{m^5}{120 \pi} \frac 1 s\nn\\
c  &\to &  c - \frac{7 m^4}{180 \pi} 
 \frac 1 {s}
\quad
d\, \to \, d-\frac{23 m^3}{630 \pi} \frac 1 s
\label{n8}\\
e &\to &e - \frac{m^2} {45 \pi}
\frac 1 {s}
\quad
f  \,\to  \, f +\frac{2m}{315 \pi} \frac 1 s\nn\\
g  &\to &  g + \frac{41}{20790 \pi} 
 \frac 1 {s} \nn
\eeq
The quantities $\alpha = \{a, b, c, d, e, f, g \}$ are a set of free parameters of
the theory to be determined experimentally. In principle we are free
to perform finite renormalizations at our choice of all the parameters. 

We find natural to perform two further renormalizations.
First it is possible to determine the asymptotic behavior of the $A_i$ for $m\to\infty$. The finite pieces not vanishing in the
limit $m\to \infty$ are all of the same type appearing in the classical 
energy. Our first finite renormalization is such that those pieces are 
cancelled. As a result, only the "quantum contributions" are finally included, 
because, physically, a quantum field of infinite mass is not expected 
to fluctuate. The resulting $A_i$ will be  called $A_i ^{(ren)}$. 

Concerning $Z$ we have constructed a numerical fit of $Z$ by a polynomial 
of the form
\beq
P(m) = \sum_{i=0}^6 c_i m^i, \nn
\eeq
and then subtracted this polynomial from $Z$. As explained above, 
this is nothing
else than an ulterior finite renormalization. The result will be 
denoted by $Z^{(ren)}$.

Summing up, we can write the complete energy as
\beq
E=E_{class} +E_0^{(ren)}, \label{egesend}
\eeq
where $E_{class}$ is defined as in  
(\ref{n1}) with the renormalized parameters $\alpha$ 
and $E_0^{(ren)}= Z^{(ren)} + \sum_{i=-1}^5 A_i^{(ren)}$. 

Figure \ref{grafico} shows the quantum contribution $E_0^{(ren)}$ as a function of the bag radius. Remarkably the Casimir contribution 
of the massive spinor is stabilizing: the energy exhibits a minimum corresponding to a stable geometrical configuration. 

In the next section we will discuss the parameters of the model and the stabilization mechanism in greater detail.

\begin{figure}[t]
\begin{center}
\includegraphics[width=12cm]{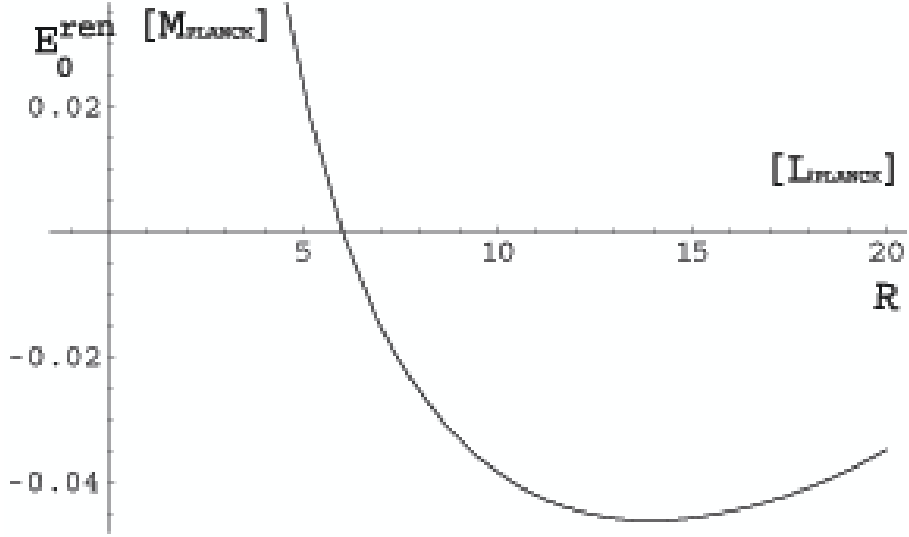}
\caption{$E_0^{(ren)}$ as a function of the bag radius R for a proper choice of parameters (in particular m=1 has been assumed). Only the contribution j=3/2 has been included as the leading one.}
\label{grafico}
\end{center}
\end{figure}

\section{Moduli stabilization and parameters space}

Is it possible to achieve a realistic model with a "natural" choice of the parameters? 
To answer this question it is necessary to remember that
our purpose is to stabilize the entire geometrical configuration of the system which amounts to stabilizing {\it both} moduli.
As we already said, it is not possible 
to introduce in a model ultralight scalar fields with a generic coupling with matter, since phenomenological
constraints must be faced. In this model the two moduli interact with matter in a known way. In fact, we could parametrize the strength of this coupling,
since we know explicitly the conformal factors (4.78), but our approach will be different: dangerous couplings 
are kept under control because we are giving a (large enough) mass
to the moduli. It seems now worthwhile to focus on the stabilization mechanism.

\subsection{R-modulus stabilization: the chameleon mechanism}

As we already mentioned in section 3, the possible interpretation of the R-modulus as a chameleon field
has been investigated in \cite{radion}. Following their approach we will add a run-away bare potential for the radion
in the form:

\begin{equation}\label{champot}
V_{add}(R)= M^4 e^{\left(\frac{M}{m_{pl} R}\right)^n}
\end{equation}
where $M=10^{-3}$ eV in order to trigger acceleration now. It is not a
quintessence potential since it does not converge to zero at
infinity. It should be considered as a very flat potential which may 
appear from some non-perturbative physics for the radion.

For the potential above, the minimum of the effective potential
\begin{equation}
V_{eff}(R) = V_{add}(R) + \rho_m (1 + \frac{\xi}{2} R^2)
\end{equation}
is given by
\begin{equation}
R_{min}^{n+2}=n\left(\frac{M}{m_{pl}}\right)^{n} \frac{V_{add}(R_{min})}{\xi \rho_m},
\end{equation}
(see formulas \ref{approx},\ref{fattori}).

For $V_{add}=O( M^4)$ this leads to a cosmological value for the radion field given by
\begin{equation}
R_{\infty}^{n+2}=O\left(\frac{M}{m_{pl}}\right)^{n}.
\end{equation}
Notice that $R_{\infty}=O(10^{-10})$ for $n=1$ implying that solar tests are
automatically satisfied.

The mass of the radion at the minimum reads
\begin{equation}\label{massarad}
\frac{m^2}{H^2}= 3\xi \Omega_m \left[ n(n+2)  + n^2 (\frac{M}{m_{pl}})^n
  \frac{1}{R^n} \right].
\end{equation}
The interesting regime is obtained for small distances, then the matter density is high and $m \gg H$, the field "sits" on 
the minimum of the effective potential (i.e. the minimum acts as an attractor): the radion is stabilized. 

Although the chameleon mechanism does work also without runaway potentials \cite{gubserk}, 
the presence of the potential \ref{champot} is crucial in our model. Had we considered only
the classical potential \ref{poti} for the radion, we would have been forced to face the following problems:
\begin{itemize}
\item It is not possible to achieve the competition between the 
potential ($V_{class} \sim Cosh(R)^{\frac{4-4 \alpha^2}{1+2 \alpha^2}}$) and the matter coupling (see formula \ref{fattori}). 
\item The value $R=0$ for the radion is not forbidden by the classical potential. 
The falling of the hidden brane into the naked singularity ($R=0$) must be avoided if our intention is to maintain 
a supergravity approximation \cite{brax1}. 
\end{itemize}

It seems noteworthy that the Q-modulus is not a chameleon 
since it is not possible to achieve the desired competition between the potential ($V_{class}(Q) \sim Q^{-\frac{3}{\beta}}$) 
and the coupling to matter ($A(Q) \sim Q^{-\frac{1}{2 \beta}}$). To stabilize the Q-modulus we suggest to exploit the 
Casimir energy of the bulk spinor evaluated in the previous section.

\subsection{Q-modulus stabilization: the Casimir energy}
The stabilization of the system is based on the presence of (A) {\it two} 
scalar degrees of freedom (the positions of the two branes are parametrized
independently) and of (B) a cosmological attractor 
that pushes the hidden brane towards the singularity.

As mentioned above, the Casimir energy of the massive spinor 
guarantees a stable geometrical configuration for the 
5D bag. The connection 
between the bag and the braneworld scenario is the conformal transformation (\ref{trconf}). 
Had we had only one scalar field in the low energy description of the model, than 
only the distance between the two branes would have been physically 
relevant (consider the radion in Randall-Sundrum two branes model) or, equivalently, 
it would have been possible to shift the zero of the fifth dimension: moving the two branes away in the bulk while keeping fixed the
distance between them. In this case the conformal transformation (\ref{trconf})
would not have mapped the two branes set up into a ball, since only the bulk singularity corresponds to the centre of the ball.

The second crucial ingredient is the presence of a cosmological 
attractor that pushes the hidden brane towards the bulk singularity \cite{palma}. The visible brane evolution
becomes independent of the hidden brane and the braneworld system is described by a {\it single}-scalar-tensor theory. 
For this reason 
we will assume a very small value for the R-modulus (the hidden brane is 
close to the singularity/bag centre) in a late-time cosmology
we are interested in. In this way we can "forget" about the R modulus (it is fixed at small values and stabilized by the chameleon
mechanism) if the 
efficiency of the attractor is high enough. The Q-modulus corresponds to the remaining scalar degree of freedom (position of the 
visible brane/bag surface) and the last step in the stabilization process will be to add the Casimir energy
as a stabilizing contribution for the bag surface. 

To proceed further, it would be necessary to study in full detail the role of the conformal transformation to the Einstein
frame (i.e. the cocycle function, for a pedagogical treatment see \cite{kirsten}) and write the full moduli potential. 
However, we will consider
a reasonable value for the Q-modulus ($Q \sim 1$) for the discussion of the neutrino mass.

\subsection{Parameters Space: Neutrino Dark Energy}

Here is a possible choice of parameters.
We will proceed stepwise: \\
1)set the fundamental scale of gravity as 
\beq
k_5^2 = M_{fund}^{-3}, \\
M_{fund} \sim M_{Planck} \sim 10^{18} GeV.
\eeq
\\
2)Choose $ \alpha^2 \lesssim 10^{-6}$
in order to face with phenomenological constraints (\cite{palma, gonz}).
\\
3)Set the tension scale and the 5D neutrino mass near the fundamental scale of gravity:\\
\beq
k \sim m \sim 10^{18} GeV.
\eeq
4) The Newton's constant as measured by 
Cavendish experiments in the visible brane is given by (see also \cite{palma}): 
 \\
\beq
\frac{Q^2}{k k_5^2 (1+2 \alpha^2) a^2(\phi)}=\frac{1}{8 \pi G}= (10^{18} GeV)^2.
\eeq
\\

For the screening potential of the form (\ref{champot})  
the mass scale has to be chosen such that $M \approx 10^{-3}$~eV if  
the radion must be a dark energy candidate. This tuning is no more
than that required by a cosmological constant.

The classical potential (\ref{poti}) shows a scale given by
\beq
V_{class} \simeq \frac{6 (T-1) k}{k_5^2} e^{\frac{-12 \alpha^2 S}{1 + 2 \alpha^2}} \simeq 6 (T-1) M_{Planck}^4.
\eeq
\\
The correct dark energy scale is recovered imposing the usual (cosmological constant) fine-tuning $T-1 \sim 10^{-120}$.\\

On cosmological distances, if we consider $n=15$ in the potential we obtain $R_{\infty} \sim 10^{-27}$. In this way the 
neutrino mass is strongly suppressed by the large separation of the branes. In fact, choosing 
\beq
m=\frac{k}{20}
\eeq
and evaluating the neutrino mass in the Einstein frame for $\bar Q \sim 1$ and $\bar R \sim 10^{-27}$ (corresponding
to the minimum of the moduli potential after the discussed stabilization), we can write
\beq
\tilde m_{\nu}= Y_5 f_0^R(\phi) v_b \frac{\sqrt{a(\phi)}}{Q} \sim 10^{-3} eV.
\eeq

Remarkably, it is possible to obtain a meV-mass for neutrinos with a reasonable branes configuration in five dimensions.
Happily, in the minimum $(\bar Q, \bar R)$ of the total potential we can write
\beq
\frac{V_{tot}^{1/4}(\bar Q,\bar R)}{\tilde m_{\nu}(\bar Q,\bar R)} = f(T, k_5, k, m, \alpha, M, n ) \sim 1,
\eeq
without fine-tuning the parameters (with the only exception of the usual cosmological constant fine-tuning, a 
proper choice of $n=15$ in the potential and a small value of $\alpha$).
With this last configuration the radion mass on cosmological distances is $m \sim 10 H$, the 
field can roll and the neutrino mass is {\it variable} with time. Also, notice that the neutrino mass is {\it space}-dependent: 
this can be a source of phenomenological consequences.


\section{Conclusions}
In this paper a braneworld model for neutrino Dark Energy has been presented. The model is a supersymmetric extension
of the Randall-Sundrum two branes model. After supersymmetry breaking the system 
becomes dynamical and a (classical) moduli potential is generated. 

In this article we have achieved the following results: \\
1) In our model a 4D neutrino mass as a function of two scalar 
degrees of freedom (moduli) has been explicitly calculated rather than inserted by-hand.

2) The moduli stabilization issue is addressed with a de Sitter geometry for the branes. The chameleon 
mechanism is exploited to stabilize the radion following the approach of \cite{radion}. As far as the stabilization of the remaining
modulus is concerned, we suggest to exploit the
Casimir energy of massive neutrinos as a stabilizing contribution.

3) A direct relationship has been established between neutrino mass scale and Dark Energy scale.
We stress again that the only dynamical behaviour associated to moduli is encoded into the chameleon field R.
Namely, on cosmological distances, the densities are tiny and the 
radion can roll on cosmological time scales, while on smaller scales, 
the chameleon acquires a large mass (radion stabilization).
The neutrino mass is {\it variable} with time on cosmological distances and possible deviations
from a pure cosmological constant behaviour can be observed in the DE sector 
(we will further discuss these issues in a future work). The neutrino mass is also space dependent and this can imply
phenomenological consequences.

It seems to us that this model suggests a stronger connection 
between DE and neutrino physics. To the best of our knowledge it is the first attempt 
of connecting neutrino mass to Dark Energy
from the standpoint of brane models. 

\acknowledgements{Special thanks are due to Antonio Masiero and Massimo Pietroni. 
They patiently guided me and they were a constant source of inspiration during the development of this article. }

\appendix

\section{D-polynomials}

In this appendix we collect the basic formulas necessary to obtain the explicit form of
the polynomials \ref{asympol}.
We start remembering the uniform asymptotic expansion of the Bessel functions $I_\nu (k)$ as $\nu, k \rightarrow \infty$ with 
$\nu /k$ fixed. One has 

\beq I_{\nu} (\nu z) \sim \frac
1 {\sqrt{2\pi \nu}}\frac{e^{\nu \eta}}{(1+z^2)^{\frac 1
4}}\left[1+\sum_{k=1}^{\infty} \frac{u_k (t)} {\nu
^k}\right], \nn
\label{eq2.9} 
\eeq 
with $t=1/\sqrt{1+z^2}$ and $\eta=\sqrt{1+z^2}+\ln [z/(1+\sqrt{1+z^2})]$.
The coefficients are determined by the recurrence formula \cite{abramo}
\beq
u_{k+1} (t) =\frac 1 2 t^2 (1-t^2) u'_k (t) +\frac 1 8
\int\limits_0^t d\tau\,\, (1-5\tau^2 ) u_k (\tau ), \nn
\label{eq2.10}
\eeq 
starting with $u_0 (t) =1$. As is clear, all the $u_k (t)$
are polynomials in $t$.

Since boundary conditions involved Bessel functions and their derivatives, we need 
also the expansion \cite{abramo}
\beq
I_{\nu} ' (\nu z )\sim
\frac 1 {\sqrt{2\pi \nu}} \frac{e^{\nu \eta} (1+z^2)^{1/4}} z
\left[1+\sum_{k=1}^{\infty}\frac{v_k(t)}{\nu^k}\right], \nn \label{eq2.48}
\eeq
with the $v_k (t)$ determined by
\beq
v_k (t) = u_k (t) +t (t^2 -1) \left[
\frac 1 2 u_{k-1} (t) +t u_{k-1} ' (t) \right]. \nn \label{eq2.49}
\eeq

Exploiting the previous formulas the relevant asymptotics can be found. With the notation
\beq
\Sigma_1 = \left[1+\sum_{k=1}^{\infty} \frac{u_k (t)} {\nu
^k}\right], \Sigma_2 = \left[1+\sum_{k=1}^{\infty} \frac{v_k (t)} {\nu
^k}\right], mR=x \nn
\eeq
it reads

\beq
&& \ln[I^2_{j}(zj)(1 + \frac{1}{z^2}-\frac{2x}{z^2 j}) + I'^2_j(zj) + \frac{2}{zj}(x-j)I_j(zj)I'_j(zj)]\simeq \nn \\
&& \simeq \ln\left\{\frac{e^{2j \eta}(1+z^2)^{1/2}}{2 \pi j z^2} [(1+\frac{1}{z^2} - \frac{2x}{z^2 j}) \frac{z^2}{1+z^2} \Sigma_1^2 + \Sigma_2^2 +
 \frac{2}{j} (x-j) \frac{1}{(1+z^2)^{1/2}} \Sigma_1 \Sigma_2]  \right\} = \nn \\
&&= \ln\left\{(\frac{e^{2j \eta}(1+z^2)^{1/2}}{2 \pi j z^2} 2(1-t))\frac{(1-\frac{2xt^2}{j}) \Sigma_1^2 + \Sigma_2^2 +
2t \frac{x-j}{j} \Sigma_1 \Sigma_2 }{2(1-t)}  \right\}. \nn
\eeq

In this way we are led to the definition of D-polynomials as follows
\beq
\sum_{k=1}^{\infty} \frac{D_k(x,t)}{j^k}=\ln\left\{\frac{1}{2(1-t)}[(1-\frac{2xt^2}{j}) \Sigma_1^2 + \Sigma_2^2 +
2t \frac{x-j}{j} \Sigma_1 \Sigma_2] \right\}. \nn
\eeq


\end{document}